\documentclass[manuscript]{aastex}

\usepackage{cases}
\usepackage{ulem}
\usepackage{color}
\usepackage{mathrsfs}

\shorttitle{Radio Properties of GRB Afterglows and their host galaxies}
\shortauthors{Zhang, Chandra, Huang \& Li}

\begin{document}

%% LaTeX will automatically break titles if they run longer than
%% one line. However, you may use \\ to force a line break if
%% you desire.

\title{The Redshift Dependence of the Radio Flux of Gamma-Ray Bursts and Their Host Galaxies}

%% Use \author, \affil, and the \and command to format
%% author and affiliation information.
%% Note that \email has replaced the old \authoremail command
%% from AASTeX v4.0. You can use \email to mark an email address
%% anywhere in the paper, not just in the front matter.
%% As in the title, use \\ to force line breaks.

\author{Z. B. Zhang\altaffilmark{1,2}, P. Chandra\altaffilmark{3}, Y. F. Huang\altaffilmark{4} and D. Li\altaffilmark{5,6}}
%\affil{Department of Physics, College of Sciences, Guizhou
%University, Huaxi district, Guiyang, Guizhou 550025, China }
%
%%\author{Y.-F. Huang\altaffilmark{2}}
%\affil{Department of Astronomy, Nanjing University, Nanjing 210093, China}
%\email{hyf@nju.edu.cn}

%\and

%\author{R. J. Hanisch\altaffilmark{5}}
%\affil{\\The Joint Research Center for Astronomy between National Astronomical
%%Observatories, CAS and Guizhou University, Guizhou, China}

%% Notice that each of these authors has alternate affiliations, which
%% are identified by the \altaffilmark after each name.  Specify alternate
%% affiliation information with \altaffiltext, with one command per each
%% affiliation.
\altaffiltext{1}{College of Physics and Engineering, Qufu Normal University, Qufu 273165, China; z-b-zhang@163.com}
\altaffiltext{2}{Department of Physics, College of Physics, Guizhou University, Guiyang 550025, China}
%\altaffiltext{3}{Department of Physics and Astronomy, University of Nevada, Las Vegas, NV 89012, USA; zhang@physics.unlv.edu}
%\altaffiltext{4}{Department of Astronomy, School of Physics, Peking University, Beijing 100871, China}
%\altaffiltext{5}{Kavli Institute for Astronomy and Astrophysics, Peking University, Beijing, 100871, China}
%\altaffiltext{6}{National Radio Astronomy Observatory, P.O. Box O, Socorro, NM 87801, USA}
\altaffiltext{3}{National Centre for Radio Astrophysics, Tata Institute of Fundamental Research, PO Box 3, Pune 411007, India}
\altaffiltext{4}{Department of Astronomy, Nanjing University, Nanjing 210023, China; hyf@nju.edu.cn}
\altaffiltext{5}{National Astronomical Observatories of China, Chinese Academy of Sciences, 20A Datun Road, Beijing 100020, China}
\altaffiltext{6}{CAS Key Laboratory of FAST, NAOC, Chinese Academy of Sciences}
%% Mark off your abstract in the ``abstract'' environment. In the manuscript
%% style, abstract will output a Received/Accepted line after the
%% title and affiliation information. No date will appear since the author
%% does not have this information. The dates will be filled in by the
%% editorial office after submission.

\begin{abstract}
 Using multiwavelength observations of radio afterglows, we confirm the hypothesis that the flux density of gamma-ray bursts (GRBs) at a fixed observing frequency is invariable when the distance of the GRBs increases, which means the detection rate will be approximately independent of redshift. We study this behavior theoretically and find that it can be well explained by the standard forward shock model involving a thin shell expanding in either a homogeneous interstellar medium (ISM) or a wind environment. We also found that short GRBs and supernova-associated GRBs, which are at relatively smaller distances, marginally match the flux-redshift relationship and they could be outliers. We rule out the assumption that the medium density evolves with redshift as $n\propto(1+z)^4$ from the current measurements of $n$ and $z$ for short and long GRBs. In addition, the possible dependence of host flux on the redshift is also investigated. We find that a similar redshift independence of the flux exists for host galaxies as well, which implies that the detection rate of radio hosts might also be independent of the redshift. It is also hinted that most radio hosts have the spectral indices ranging from $\beta_h\simeq-1$ to 2.5 in statistics. Finally, we predict the detection rates of radio afterglows by the next-generation radio telescopes such as the Five-hundred meter Aperture Spherical Telescope (FAST) and the Square Kilometer Array (SKA).
\end{abstract}
%FAST in particular is outstanding in detecting high redshift radio afterglows.

%% Keywords should appear after the \end{abstract} command. The uncommented
%% example has been keyed in ApJ style. See the instructions to authors
%% for the journal to which you are submitting your paper to determine
%% what keyword punctuation is appropriate.


%% From the front matter, we move on to the body of the paper.
%% In the first two sections, notice the use of the natbib \citep
%% and \citet commands to identify citations.  The citations are
%% tied to the reference list via symbolic KEYs. The KEY corresponds
%% to the KEY in the \bibitem in the reference list below. We have
%% chosen the first three characters of the first author's name plus
%% the last two numeral of the year of publication as our KEY for
%% each reference.

%% Authors who wish to have the most important objects in their paper
%% linked in the electronic edition to a data center may do so by tagging
%% their objects with \objectname{} or \object{}.  Each macro takes the
%% object name as its required argument. The optional, square-bracket
%% argument should be used in cases where the data center identification
%% differs from what is to be printed in the paper.  The text appearing
%% in curly braces is what will appear in print in the published paper.
%% If the object name is recognized by the data centers, it will be linked
%% in the electronic edition to the object data available at the data centers
%%
%% Note that for sources with brackets in their names, e.g. [WEG2004] 14h-090,
%% the brackets must be escaped with backslashes when used in the first
%% square-bracket argument, for instance, \object[\[WEG2004\] 14h-090]{90}).
%%  Otherwise, LaTeX will issue an error.

\section{Introduction}

The radio afterglow of gamma-ray bursts (GRBs) was first discovered by Frail et al. (1997)
for GRB 970508. Long-lasting radio afterglows are essentially immune to the geometry of
the initial ejecta and thus can offer us an ideal way to estimate the true energy $E_\gamma$,
this is because the radio afterglows are emitted at relatively later epochs when the Lorentz
factor drops to sub-/non-relativistic levels (Berger, Kulkarni \& Frail 2004; Shivvers \&
Berger 2011; Wygod, Waxman \& Frail 2011; Mesler \& Pihlstr\"{o}m 2013). There are some additional advantages of radio
observations, such as: (1) comparing with X-ray and optical emissions, the radio
afterglow lasts much longer that more detailed observations can be performed and can provide
key clues to diagnose the intrinsic properties of the explosion; (2) radio observations can
play an important role in revealing the structure of surrounding medium, the geometry of
the outflow (i.e. measuring the tiny angular size of afterglows via interstellar scintillation),
as well as in revealing the progenitors of the explosions (e.g. Frail 2003); (3) like many other astronomical
objects such as compact stars, supernova (SN) remnants, interstellar medium, intergalactic medium,
and radio lobes and jets of galaxies driven by central black holes, GRBs produce synchrotron
radio emissions with a ``steep'' spectrum at later epochs, which indicates that their intensities
increase strongly toward the low-frequency regime, thus they can be more conveniently observed
in radio for a relatively long period. It is interesting to note that far-infrared observations
show that the detection rate of GRB hosts is consistent with the idea
that GRBs trace the cosmic star formation rates (Kohn et al. 2015).
%(4) Measuring the polarised radio waves at low frequencies will offer us a new window to study cosmic magnetism. Low-frequency radio afterglow, on the other hand, is emitted by electrons with lower energies which suffer less from energy losses and hence can propagate farther away from their origins into regions with weak magnetic fields.

Ciardi \& Loeb (2000) argued that the detectability of radio afterglows by ground-based radio
telescopes is somewhat independent of redshifts. It is mainly based on theoretical studies
showing that the dependence of the radio brightness on the redshift becomes increasingly weaker
at higher redshifts (Ciardi \& Loeb 2000; Gou et al. 2004). This argument has been proved by
Karl G. Jansky Very Large Array (JVLA) observations and the Expanded Very Large Array
Project (EVLA) at 8.5 GHz directly (Frail et al. 2006; Chandra \& Frail 2012). In addition, Chandra \& Frail (2012) showed that the detection rate starts to become independent of redshift after a redshift of 3. Such an effect
makes it possible for us to observe very distant GRBs (up to $z>15$) with large radio
telescopes (e.g. Zhang et al. 2015). However, how the radio fluxes of
GRB host galaxies evolve with their redshifts is still largely uncertain.

Observationally, roughly one-third of all GRBs with precise localization have
been detected at radio frequencies (Chandra \& Frail 2012; Chandra 2016). This rate is much lower
than those at higher observing frequencies, where for instance ~93\% of GRBs observed in gamma-rays
are also detected in X-ray bands and ~75\% are detected in optical bands. Furthermore,
radio afterglows are more difficult to detect at lower radio frequencies owing to the
self-absorption or influence of the host galaxies (e.g. Berger, Kulkarni \& Frail 2001; Berger 2014;
Li et al. 2015). Chandra \& Frail (2012) presented a large radio afterglow sample of 304 GRBs,
including 33 short-hard bursts, 19 X-ray flashes, and 26 GRB/SN candidates. Their sample also
includes several low-luminosity bursts and high-redshift bursts, whose radio afterglows are even
more difficult to detect due to their low energetics or large distances, and the interference from the
host galaxies. Recently, Li et al. (2015) proposed an interesting method to infer the contributions
of the host galaxies at observational frequencies of $\nu\leq$ 10 GHz.
They found that at lower radio frequencies, the contribution of hosts becomes more important.
An empirical relation was derived to approximate the frequency dependence of the host contribution,
which can help to significantly increase the detectability of radio afterglows and should be
particularly helpful in the upcoming era of large telescopes (Burlon et al. 2015; Zhang et al. 2015).

The properties of GRB host galaxies are important in understanding the nature of GRBs. For instance, one can use the hosts to study the large-scale environments, the burst energetics (once the redshift is determined from optical spectrum of the host galaxy), and further constraints on the nature of GRB progenitors.
Berger (2014) pointed out that different populations of short and long GRBs also
differ significantly in their host galaxies (see also Zhang et al. 2009).
Savaglio et al. (2009) have used optical and near IR (NIR) photometry and spectroscopy methods to study
stellar masses, star formation rates, dust extinctions, and metallicities of
a large set of GRB hosts. They found that GRBs can be used as a good probe to study star-forming galaxies.
Their samples include 46 objects ranging in a redshift interval of $0 < z < 6.3$
with an average of $z\sim1$. In their data set, about 90\% of the hosts have relatively small redshifts
of $z < 1.6$. Stanway et al. (2014) later reported their radio continuum observations of 17 GRB host galaxies
with the Australia Telescope Compact Array (ATCA) and VLA at 5.5 and 9.0 GHz, respectively. Their samples span in a redshift
range of 0.5 -- 1.4.
% the following sentence should be deleted? *********************
% They speculated that the previously reported radio afterglow of GRB 100621A (Greiner et al., 2013)
% could be contributed by a host galaxy flux.
% the above deleted?  ********************
Recently, Kohn et al. (2015) presented their analysis of the far-infrared properties of an ``unbiased''
set of GRB host galaxies. Their samples include 20 \textit{BeppoSAX} and \textit{Swift} GRBs, among which
eight bursts are listed with known redshifts (the average value is $z=3.1$). They constrained the dust
masses and star formation rates (SFRs) of the hosts, and found that GRBs may trace the SFR of luminous galaxies in an unbiased way up to $z>2$. The interesting result by Li et al. (2015) that the ratio of the host flux density to the peak flux of GRB
afterglow is tightly correlated with the observing frequency may also shed new light on the environment
properties of GRBs. However, we notice that little is known about the spectra of GRB hosts in radio bands
except for the special event of GRB 980703 (Berger, Kulkarni \& Frail 2001), whose host spectral
index was estimated as $\beta_h\thickapprox-1/3$ from three data points at different frequencies. Observationally,
most normal galaxies, such as M82 and our Milk Way Galaxy, usually have the spectral power-law index of $-3/4$
(Condon 1992; Carilli \& Yun 1999). In principle, the synchrotron radiation mechanism may
result in a positive spectrum index in the radio bands (e.g. Sari et al.1998; Gao et al. 2013). The positive indices
observed in a few GRBs thus indicate that they could be originated from some special types of galaxies, such as starburst or active galaxies.

In this study, we present a large data set for GRBs whose afterglows as well as their hosts are observed in
radio wavelengths. The data are collected from the literature and are described in Section 2.
In Section 3, we re-examine whether the radio fluxes are dependent on the redshifts with multiple-band
observational data of GRB afterglows, and compare the results with theoretical predictions.
We also examine how the radio fluxes of the hosts evolve with the redshifts from the data set.
The detectability of GRBs by different large radio
telescopes, such as the Square Kilometer Array (SKA, Dewdney et al. 2009) and the Five-hundred-meter Aperture
Spherical radio Telescope (FAST, Nan et al. 2011; Li et al. 2013) are studied.
Finally, we present our conclusions and brief discussion in Section 4.

\section{Data Collection}

For the purpose of studying the flux-redshift dependence of radio afterglows,
17, 30 and 54 GRBs are available in  Chandra \& Frail (2012) at three frequencies
of $\nu$ =1.43, 4.86 and 8.46 GHz, respectively. They were all measured with peak
radio fluxes, peak times and redshifts. We will use these observational data in our
current study. Note that two short GRBs (050724 and 051221) and three SNe-associated
GRBs (980425, 031203 and 060218) are included in these samples. Although the numbers of
these special GRBs are too limited, they might still be helpful in hinting us the
systematic differences between them and normal long GRBs.

In general, the radio hosts of GRBs are so faint that only about three hosts could
be detected each year by all current ground-based radio telescopes.
However, it is interesting to investigate the flux-redshift dependence of GRB host
galaxies in radio bands and compare it with that of afterglows. For this target, we have
also collected 37 long bursts with 47 measured host flux densities at several low/medium
frequencies of 1.43, 3.0, 4.9, 5.5, 9.0, 37.5 and 222 GHz. The sample selection
criteria are as follows: (1) the radio afterglow of the corresponding GRB was observed;
(2) the redshift was measured; (3) the host flux densities had been reported in the literature.
The data and their references are listed in Table 1. In this table, Columns (1)-(8)
correspond to the burst names, durations ($T_{90}$), cosmological redshifts, isotropic
$\gamma$-ray energies, observing frequencies, host flux densities ($F_{host}$), references of $F_{host}$, and telescopes,  respectively.

In Table 1, the first set of entries ($N=16$, i.e., from Line 1 to Line 24)
represents relatively bright events of 24 measurements with the peak of the radio afterglow being clearly observed as well.
For these events, we are assured that the host fluxes have been relatively accurately measured and the
interference from their afterglows have been subtracted. Hereafter, we call this sub-sample as the
``Gold-Host Sample''. The second data set ($N=18$, i.e., from Line 25 to Line 46) in Table 1 denotes
those 18 hosts with 22 measurements without observed peak fluxes of radio afterglows. In Table 1, there are 11 and 16 host measurements
collected from Perley et al. (2015) and Stanway et al. (2014), respectively, and around one third GRBs are associated with Supernovae.

As demonstrated in some previous works, GRB 100418A is an ultra long burst without supernova association (e.g. Jia et al. 2012; Niino et al. 2012). Its radio flux densities reached $363\pm48$ $\mu$Jy and $199\pm57$ $\mu$Jy at 5.5 and 9.0 GHz, respectively.
The unusually high radio flux densities are believed to be from the radio afterglows instead of the host galaxies (Stanway et al. 2014). In addition, two high-redshift GRBs (050904 and 090323) are not included in Table 1.
The peak fluxes of radio afterglows at 8.46 GHz are measured for these three GRBs, but the host fluxes at
our interested frequencies are not directly available. For example, the host galaxy
of GRB 090423 was not detected at higher frequencies of $\nu_{obs}$=222 and 37.5 GHz by the Atacama Large Millimeter Array (ALMA) and the ATCA,
but note that the upper limits of the host flux at these frequencies have been
constrained by Berger et al. (2014) and Stanway et al. (2011).
%\textcolor{red}{\sout{For these hosts of GRBs 050904, 090323, and 090423, one can adopt the so-called Radio Ratio of Flux (RRF) method (Li et al. 2015) together with error propagation to estimate their host flux densities at 8.46 GHz. The required radio peak flux densities of the three high-redshift bursts had been reported in Chandra \& Frail (2012).}}

\section{Results}

In this section, we will firstly study the redshift dependence of the radio flux densities of GRB afterglows and
hosts based on our samples. Then we use the newly-found flux-redshift dependence to constrain the spectral
parameters of these host galaxies. Finally, we investigate the detectability of GRB radio
afterglows by the next-generation radio instruments, such as the Low-Frequency Array or LOFAR (van Haarlem et al. 2013),
FAST (Nan et al. 2011; Li et al. 2013) and SKA (Dewdney et al. 2009), etc. In the following theoretical
calculations, we take typical values for the key parameters of the forward shock model. For example,
the microphysical parameters of electrons and magnetic fields are $\varepsilon_{e}=0.1$ and
$\varepsilon_B=0.01$ (e.g. Panaitescu \& Kumar 2002; Zhang, Kobayashi \& M\'{e}sz\'{a}ros 2003),
respectively. The average isotropic energy of our radio-selected sample with peak flux measurements
from Chandra \& Frail (2012) is about $E_{iso}=2\times10^{52}$ erg, thus this value will be
used in our numerical calculations below. Finally, the power-law index of electron distribution is assumed to be $p=2.3$.
% However, physical origin of the first component in the radio afterglows is still amphibolous,
% although some authors have suggested reverse shock contributions (Chandra \& Frail 2012;
% Resmi \& Zhang 2015) or multiple forward shocks overlapped (Li, Zhang \& Rice 2015).
% We only consider the radio emissions generated from the forward shocks below

\subsection{Flux-Redshift Dependence for GRB Afterglows}

The peak flux densities of radio afterglows at $\nu=$ 1.43, 4.86 and 8.46 GHz are plotted against
the redshifts in Figures 1 --- 5. It can be easily seen that the radio flux density does exhibit a weak
dependence on the redshift. Generally speaking, the peak flux densities are weaker for more
distant events. In fact, such a weak dependence has been noticed in several previous studies
(e.g., Ciardi \& Loeb 2000; Gou et al. 2004; Frail et al. 2006; Chandra \& Frail 2012).
Below, we give a quantitative explanation for the dependence in the frame work of the
standard forward shock model.

\subsubsection{Methodology}
Systematical analytical solutions for GRB afterglows involving forward shock emission in either
the fast cooling regime or the slow cooling regime have been addressed by many authors (e.g.,
M\'{e}sz\'{a}ros \& Rees 1997; Sari et al. 1998; Chevalier \& Li 1999; Huang et al.1999, 2000;
Zhang \& M\'{e}sz\'{a}ros 2004; Wu et al. 2005; Zhang et al. 2006; Gao et al. 2013).
Following usual treatments, we assume that the ambient density at radius $R$ is $n=AR^{-k}$ cm$^{-3}$,
where $k$ is a constant index characterizing the density form of the medium and $R$ is the
radius of a blast wave in unit of centimeter. There are mainly
two kinds of density forms. In the homogeneous interstellar medium (ISM) case, the density is a constant
and we have $k=0$. In the stellar wind case, the density decreases outward so that we have
$k=2$. In the latter case, we can further write the density as $n=AR^{-2}$ cm$^{-3}$,
where $A=3\times10^{35}A_{*}$ cm$^{-1}$ (a typical wind parameter of $A_{*}=0.2$ will be
taken in our calculations, see below). We assume $n_0=0.1$ cm$^{-3}$ and $A_{*}\simeq0.2$ to be the best parameters for the ISM and wind cases, respectively. Wu et al. (2005) argued that the parameter $A_{*}$ should be quite small
to fit typical observational data. In fact, the deduced values of $A_{*}$ for a few long GRBs
span four orders of magnitude, ranging from $5\times10^{-4}$ to 3.5, with the median value
being 0.18 (Panaitescu \& Kumar 2002; Price et al.,2002; Dai \& Wu 2003; Panaitescu \& Kumar 2003;
Chevalier, Li \& Fransson 2004). Therefore, our wind parameter of $A_{*}\simeq0.2$ is a reasonable value.
However, we should bare in mind that although we have taken a typical set of parameters to calculate
the afterglow light curves, these parameters actually could differ from burst to burst, thus the
actual afterglow light curves should also vary accordingly. In Figures 1 -- 5, we have also varied
several microphysical parameters to illustrate their effects.
Finally, we assume that the outflows are adiabatic in our calculations, since radio afterglows are usually observed at relatively late
stages. It is consistent with the fact that the radiation efficiency $\varepsilon$ is negligible after
the blast wave enters the self-similar deceleration phase.

The observed flux density at a certain frequency $\nu$ can be given after considering three
characteristic frequencies (i.e., $\nu_c$: the cooling frequency; $\nu_m$: the typical synchrotron
frequency; $\nu_a$: the self-absorption frequency) by

\begin{eqnarray}
F_{\nu}=F_{\nu,max} \times
\left\{
\begin{array}{lll}
(\nu/\nu_a)^2(\nu_a/\nu_m)^{1/3}, \ \ \ \ \ \ \ \ \ \ \nu<\nu_a;       \\
(\nu/\nu_m)^{1/3}, \ \ \ \ \ \ \ \ \ \ \ \ \ \ \ \ \ \ \ \ \nu_a<\nu<\nu_m;    \\
(\nu/\nu_m)^{-(p-1)/2}, \ \ \ \ \ \ \ \ \ \ \ \ \ \ \ \nu_m<\nu<\nu_c; \\
(\nu/\nu_c)^{-p/2}(\nu_c/\nu_m)^{-(p-1)/2}, \ \nu_c<\nu,
\end{array}
\right.
\end{eqnarray}
when $\nu_a<\nu_m<\nu_c$ (Case I) or by
\begin{eqnarray}
F_{\nu}=F_{\nu,max} \times
\left\{
\begin{array}{lll}
(\nu_m/\nu_a)^{(p+4)/2}(\nu/\nu_m)^{2}, \ \ \ \ \ \ \ \ \ \nu<\nu_m;       \\
(\nu_a/\nu_m)^{-(p-1)/2}(\nu/\nu_a)^{5/2}, \ \ \  \ \ \ \nu_m<\nu<\nu_a;    \\
(\nu/\nu_m)^{-(p-1)/2}, \ \ \ \ \ \ \ \ \ \ \ \ \ \ \ \ \ \nu_a<\nu<\nu_c; \\
(\nu/\nu_c)^{-p/2}(\nu_c/\nu_m)^{-(p-1)/2}, \ \ \ \nu_c<\nu,
\end{array}
\right.
\end{eqnarray}
when $\nu_m<\nu_a<\nu_c$ (Case II) in the slow cooling regime during the late afterglow stage.
Here the quantity $F_{\nu, max}$ denotes the flux density at the characteristic frequency of
$\nu_m$. The majority of electrons are emitting electromagnetic waves at around this frequency.
The observed flux density can peak either at $\nu_m$ or $\nu_a$ in the above two cases, thus can
be calculated as
\begin{eqnarray}
F_{\nu, t_p}(z)=F_{\nu,max}
\left\{
\begin{array}{lll}
1,\ \ \ \ \ \ \ \ \ \ \ \ \ \ \ \ (\nu_a<\nu_m<\nu_c);      \\
(\nu_m/\nu_a)^{(p-1)/2}, \ (\nu_m<\nu_a<\nu_c),
\end{array}
\right.
\end{eqnarray}
where $t_p (\equiv t_{p,obs})$ represents the observed peak time of the radio afterglow. We caution
that in each case the peak flux density will evolve into the same form as $F_{\nu, t_p}(z)=F_{\nu,max}$
eventually. In practice, there are even other three possible $\nu_a>\nu_c$ cases, requiring quite different
electron distributions (Gao et al. 2013). Those cases occur only in very rare scenarios, and are neglected in
this study.

Our recent investigations show that the radio afterglows at lower frequencies of a few GHz usually peak
at dozens of days after the bursts (Zhang et al. 2015). These peaks often occur during the Phase 3 defined
in Gao et al. (2013). If the sideways expansion effect of the jet is negligible, one can easily get the
dependence of the peak flux density on the redshift at any given frequency.
In the ISM case ($k=0$) with a constant density of $n_0$, we have $\nu_m\propto(1+z)^{1/2}$,
$\nu_c\propto(1+z)^{-1/2}$ and $F_{\nu,max}\propto(1+z)D_{l}^{-2}(z)$. While in the wind medium case ($k=2$),
we can get $\nu_m\propto(1+z)^{1/2}$, $\nu_c\propto(1+z)^{-3/2}$ and $F_{\nu,max}\propto(1+z)^{3/2}D_{l}^{-2}(z)$.
Here $D_{l}(z)$ denotes the luminosity distance given by $D_l(z)=\frac{(1+z)c}{H_0}\int_0^z\frac{dz'}{E(z')}$, where $E(z')=\frac{H(z')}{H_0}=[\Omega_m(1+z')^3+\Omega_k(1+z')^2+\Omega_\Lambda f(z')]^{1/2}$ with $\Omega_\Lambda=0.68$,
$\Omega_m=0.32$, $\Omega_k=0$, $H_0\simeq67$ km s$^{-1}$Mpc$^{-1}$ according to the latest cosmology
observations (Planck Collaboration, Ade et al. 2014),
and $f(z)=exp[3\int_0^z\frac{1+w(z')}{1+z'}dz']\equiv1$ as $w(z)\simeq-1$ for a flat $\Lambda$CDM cosmological model.
%The flux redshift dependence during the fast cooling case ($\nu_a<\nu_c<\nu_m$) can be written as
%\begin{eqnarray}
%F_{\nu, t_p}(z)\propto D_{l}^{-2}(z)
%\left\{
%\begin{array}{lll}
%(1+z)^2, \ \ \ \ \ \ \ \ \nu<\nu_a;       \\
%(1+z)^{7/6}, \ \ \ \ \ \ \nu_a<\nu<\nu_c;   \\
%(1+z)^{3/4}, \ \ \ \ \ \nu_c<\nu<\nu_m; \\
%(1+z)^{(p+2)/4}, \ \nu_m<\nu;
%\end{array}
%\right.
%\end{eqnarray}
%for the ISM medium or
%
%\begin{eqnarray}
%F_{\nu, t_p}(z)\propto D_{l}^{-2}(z)
%\left\{
%\begin{array}{lll}
%(1+z), \ \ \ \ \ \ \ \ \nu<\nu_a;       \\
%(1+z)^{2}, \ \ \ \ \ \ \nu_a<\nu<\nu_c;   \\
%(1+z)^{3/4}, \ \ \ \ \ \nu_c<\nu<\nu_m; \\
%(1+z)^{(p+2)/4}, \ \nu_m<\nu;
%\end{array}
%\right.
%\end{eqnarray}
%for the stellar WIND medium,

According to Eq. (3), in the late slow cooling phase ($\nu_a<\nu_m<\nu_c$), the flux-redshift dependence can be characterized by
\begin{eqnarray}
F_{\nu, t_p}(z)\propto(1+z)D_{l}^{-2}(z)
\end{eqnarray}
for the ISM medium, or
%as $\nu_a\propto(1+z)^{-1}$(redshift relation)
\begin{eqnarray}
F_{\nu, t_p}(z)\propto(1+z)^{3/2}D_{l}^{-2}(z)
\end{eqnarray}
in the stellar wind case. It is noticeable that both Eqs. (4) and (5) are independent of $\nu_a$.
%as $\nu_a\propto(1+z)^{-2/5}$ (redshift relation)
Instead, if the condition of $\nu_m<\nu_a<\nu_c$ is satisfied for the other slow cooling case
in Eq. (3), the peak flux-redshift dependence can be characterized by
\begin{eqnarray}
F_{\nu, t_p}(z)\propto(1+z)^{\frac{7p+3}{2(p+4)}}D_{l}^{-2}(z)
\end{eqnarray}
as $\nu_a\propto(1+z)^{(p-6)/[2(p+4)]}$ for the ISM medium, or
\begin{eqnarray}
F_{\nu, t_p}(z)\propto(1+z)^{\frac{6p+9}{2(p+4)}}D_{l}^{-2}(z)
\end{eqnarray}
as $\nu_a\propto(1+z)^{(p-2)/[2(p+4)]}$ in the wind case.

Note that all the above flux-redshift relations have been obtained on condition that the medium
density is independent of the cosmological redshift. In the constant density ISM case,
if the medium has a redshift dependence such as $n=n_0(1+z)^4$ (Ciardi \& Loeb 2000), then we can
obtain $\nu_m\propto(1+z)^{1/2}$, $\nu_c\propto(1+z)^{-9/2}$, and $F_{\nu,max}\propto(1+z)^3 D_{l}^{-2}(z)$.
In this case, our Eqs. (4) and (6) will change to
%\begin{eqnarray}
%F_{\nu, t_p}(z)\propto D_{l}^{-2}(z)
%\left\{
%\begin{array}{lll}
%(1+z)^{7/2}, \ \ \ \ \ \ \ \ \nu<\nu_a;       \\
%(1+z)^{9/2}, \ \ \ \ \ \ \nu_a<\nu<\nu_c;   \\
%(1+z)^{3/4}, \ \ \ \ \ \nu_c<\nu<\nu_m; \\
%(1+z)^{(p+2)/4}, \ \nu_m<\nu;
%\end{array}
%\right.
%\end{eqnarray}
%for $\nu_a<\nu_c<\nu_m$, and
\begin{eqnarray}
F_{\nu, t_p}(z)\propto(1+z)^3D_{l}^{-2}(z)
\end{eqnarray}
as $\nu_a\propto(1+z)^{7/5}$ for $\nu_a<\nu_m<\nu_c$, and
\begin{eqnarray}
F_{\nu, t_p}(z)\propto(1+z)^{\frac{3p+27}{2(p+4)}}D_{l}^{-2}(z)
\end{eqnarray}
as $\nu_a\propto(1+z)^{(p+10)/[2(p+4)]}$ for $\nu_m<\nu_a<\nu_c$. The peak radio luminosity
can be determined by $L_{\nu,t_p}(z)=4\pi D_{l}^2(z) F_{\nu,t_p}(z)(1+z)^{-1}$ without the k-correction,
or $L_{\nu,t_p}(z)=4\pi D_{l}^2(z) F_{\nu,t_p}(z)(1+z)^{-1}k$ with a k-correction factor of
$k=(1+z)^{\alpha-\beta}$, where $\alpha\sim0$ and $\beta\sim1/3$ are normal temporal and spectral
indices defined in $F_\nu(t) \propto t^{\alpha}\nu^{\beta}$ (Soderberg et al. 2004; Frail et al. 2006; Chandra \& Frail 2012).

\subsubsection{Model Testing}
Taking the above medium parameters ($E_{iso}$, $n_0$, $A_1$, $\varepsilon_e$ and $\varepsilon_B$),
but allowing them to vary within an order of magnitude separately, we have calculated the
evolution profiles of peak flux density versus redshift. The results are shown in Figures 1 --- 5.
From Figure 1 we see that at high frequency bands, the radio afterglows can still be largely observable
at high redshifts.
On the contrary, short and SNe-associated GRBs are more likely detected mainly in the nearby universe.
Additionally, we stress that both the ISM and the wind environment models can account for
the flux-redshift dependence. The power law index $\tau$ in the relation of $F_{\nu, t_p}\propto(1+z)^{\tau}D_l^{-2}(z)$
from Eqs. (4) --- (9) has been compared for the  three different medium cases in Table 2. Interestingly, we find that the peak fluxes drop sharply in the ISM case (with a constant density at all redshifts),
but decrease slowly in the ISM case of $n\propto(1+z)^4$. In view of the currently available observational
results in Figures 1 --- 5, we emphasize that the latter fourth power law case can be excluded empirically.
This point can be further examined below when we vary other four parameters ($n_0$, $A_{*}$, $\varepsilon_B$
and $\varepsilon_{e}$) individually for one order of magnitude to investigate the dependency of the
peak flux density on the redshift. It proves that the four parameters can independently influence the
flux-redshift evolution in a sense as shown from Figures 2 to 5. However, it is hard to judge from
Figures 1 to 5 which interstellar medium model is better in statistics.
Note that the peak flux-redshift dependence is affected not
only by the circum-burst medium structure (ISM or wind), but also by the different microphysical parameters,
such as $E_{iso}$, $n$, $A_{*}$, $\varepsilon_{e}$ and $\varepsilon_{B}$.
For a given medium structure, the variations of the microphysical parameters may influence
the peak flux-redshift relationship.
Interestingly, our theoretical investigations on the flux-redshift relation may give an upper limit
for the electron equipartition parameter as $\varepsilon_{e}\leq0.1$. It is less than the usually
assumed value of 1/3 for fast cooling electrons at early times (see Wu et al. 2005).

Theoretically, Gou et al. (2004) have studied how the medium density changes with redshift
in the framework of the forward and reverse shock model. They found that there is no correlation
between $n$ and $z$. Now we examine this issue from the observational viewpoint.
We use the medium density data derived for a number of GRBs by Chandra \& Frail
(2012) and Fong et al (2015). Particularly, Fong et al (2015) presented the medium densities for
38 short GRBs and found that most of these GRBs occurred in lower density medium ($n < 1$ cm$^{-3}$).
In Figure 6, we plot the number density versus the redshift
for these events, which include 4 short and 24 long GRBs that have both the redshift measurements
and the density estimation. This figure generally shows that the derived medium density does NOT
evolve with the redshift. In Figure 6, we specially examined the power-law
relation of $n=n_0(1+z)^4$. We take $n_0$=0.1, 1, and 10 cm$^{-3}$, and plot the
curves respectively. We see that the observational data points do not follow these curves.
Figure 6 thus clearly confirms that the number density and the redshift are not correlated with each other.

Note that the observed peak time of radio afterglows may suffer from the cosmological time dilation.
It is interesting to examine whether this effect exists in the observational data. Figure 7 shows the peak time
versus the cosmological redshift for ten GRBs with measured radio fluxes of both afterglows and hosts from
Table 1. The peak times of these GRBs were derived by Chandra \& Frail (2012).
In Panel (a), it can be clearly seen that the observed peak time
does have a tight correlation with the redshift. The best fitted relation is $t_{p,obs} \propto (1+z)$,
with a correlation coefficient of $r\simeq0.85$, which corresponds to a 99\% confidence level (not including GRB 100418A).
In Panel (b), after correcting for the cosmological time dilation effect, we see that the intrinsic peak
time is largely independent of the redshift and it tends to be a constant of about 5 days especially
at high redshifts. In both panels, GRB 100418A specially stands out as an obvious outlier. In fact,
GRB 100418A is a unique long burst without a supernova association (Niino et al. 2012). In addition to
the very late peak time of radio emission, it also has an unusual long-lasting X-ray and optical afterglow,
especially with a long optical plateau (Marshall et al. 2011). It has been suggested that this GRB can
be specially powered by continual activities of the central engine (Moin et al 2013; Li, Zhang \& Rice 2015).

\subsection{Flux-Redshift Dependence for Host Galaxies}

Now we use 36 GRB hosts (except those upper limits) listed in Table 1 to study how the host flux $F_{\nu,h}$ evolves
with the redshift. The results are plotted in Figure 8.
For this purpose, a power-law form of $F_{\nu,h}\propto \nu^{\beta_h}$ has been assumed
for the GRB hosts. As discussed in Section 3.1, the spectral luminosity of host galaxies would
similarly satisfy $L_{\nu,h}(z)=4\pi D_{l}^2(z) F_{\nu,h}(z)(1+z)^{-1-\beta_h}$,
which gives $F_{\nu,h}(z)=[L_{\nu,h}(z)/4\pi D_{l}^2(z)](1+z)^{1+\beta_h}$. If the GRB hosts can
also be regarded as a standard candle, which means their $L_{\nu,h}(z)$ concentrate in a relatively
narrow range, one can then derive the correlation between the host flux and the redshift. In fact,
the radio spectral luminosities of GRB host galaxies do concentrate at
around $\overline{L_{\nu,h}(z)}\simeq 3.6\times10^{29}$ erg s$^{-1}$Hz$^{-1}$,
as shown in the right panel of Figure 9. Optimistically, from Figure 8, we find that the GRB hosts exhibit weak flux-redshift
dependence when the distance of GRBs becomes farther and farther.
It is also found that the spectral index $\beta_h$ of hosts generally varies between -1 and 2.5,
when the three low-redshift SN-associated bursts (980425, 031203 and 060218, which seem to be
obvious outliers in Figure 8) are not included.
It does not conflict with previous results on $\beta_h$, such as $\beta_h=-0.75$ reported by Condon (1992).
The advantage of our method is that it can be used to constrain the spectral index of $\beta_{h}$ roughly when
the spectrum of the host is available but for the limited data points observationally.
% This is because the spectrum below the self-absorption frequency depends on the electrono distribution,
% one can obtain the standard index 2.5 when the synchrotron frequency is within the self-absorption
% range, that is to say, from an extragalactic non-thermal opaque synchrotron sources.
A high index of $\beta_h \simeq 2 (2.5)$ indicates that the radio emission of the GRB hosts may be
affected by synchrotron self-absorption, similar to that of GRB afterglows in the slow cooling
phase (e.g. M\'{e}sz\'{a}ros \& Rees, 1993; Paczynski \& Rhoads, 1993; Katz \& Piran, 1997).
Alternatively, the value of $\beta_h$ can also be explained by the synchrotron radiation itself
as shown in Eqs. (1) --- (2), where the host spectra will peak at $\nu_m$ ($\nu_a$) and $\beta_h$
is equal to 2 (2.5) if $\nu_a<\nu_m$ ($\nu_m<\nu_a$) is satisfied. %Interestingly, Figure 8 shows that
%higher observation frequency is not necessarily good for detecting high redshift radio hosts, which is different from radio
%afterglows (Zhang et al. 2015 and in this work).For comparison, we also
%plot those host detections with upper limits and lower Signal-Noise Ratios ($<3\sigma$) on the bottom
%panel of Figure 8, where we find that most data points are located at a region with $\beta_h<2$, of which the flux densities of faint radio hosts might be biased by the threshold of the current radio telescopes or confused with other kinds of sky noises.
It is noticeable that the majority of the fainter hosts in Figure 8 are reported by Perley et al. (2015). Unfortunately, only half of
the GRBs associated with these faint radio hosts were detected with radio afterglows. What
makes things even worse is that the peak flux measurements are unavailable for almost all of
them, except for GRB 060218. This is consistent with the fact that the radio hosts are on average
at least one order of magnitude weaker than the peak brightness of the radio afterglow.
The median flux densities at 3 and 8.5 GHz in Table 1 (excluding those upper limits)
are about $ 9.1\pm3.2$ and $23\pm9$ $\mu$Jy, respectively.

% It not only tells us a power law relation of $F_{\nu,h}\propto\nu^{2}$ for the brighter hosts
% but also hints a softer spectral index of $\beta_h\sim1/3$ for some faint hosts. The physical
% origin of the index of 1/3 is not clear yet. We speculate that these faint radio hosts could
% dominate the afterglow components in the late epoch. Interestingly, the rough relation of
% $F_{\nu,h}\propto\nu^{1/3}$ for the latter faint case is consistent with that of late radio
% afterglows, which means the radiation mechanisms of radio afterglows and hosts might be
% similar. This would be very helpful to constrain the galactic types, radiation mechanisms
% and physical origins of GRB hosts.
%On the other hand, we notice two frequency observations distribute individually and match different dependency relationship.
%It seems to indicate that two different radio hosts might be expected and verified with
%larger observational samples, especially in the larger redshift end of low radio frequencies.
%This is somewhat consistent with a previous speculation that two populations of GRBs, namely
%radio-bright and radio-faint GRBs could exist (Hancock et al. 2013).

In Figure 9, we investigate the correlation between the radio luminosity of GRB hosts and
the redshift. The average spectral luminosity of the 36 well detected GRB host galaxies in
Table 1 is $\sim3.6 \times 10^{29} $ erg s$^{-1}$Hz$^{-1}$, with a standard deviation of
$\sigma_{logL_{\nu, h}} \simeq 0.94$. When the three SNe-associated GRBs 980425, 031203 and 060218
are excluded, we can get the mean spectral luminosity as $\sim0.95\times10^{30} $erg s$^{-1}$Hz$^{-1}$.
To compare with the detection limit of FAST and SKA, we use Eq. (9) of Zhang et al (2015) to
calculate the 5$\sigma$ level sensitivities of these instruments at a representative frequency
of 1.43 GHz. A factor of $1/(1+z)$ for the cosmological time dilation effect has been considered in
the calculations. Identifying GRB host fluxes at very high redshift is a huge challenge at lower
frequencies. Even at higher frequency, as of Aug 2015, only one upper limit of the host flux density had been
obtained for high-redshift bursts (i.e. GRB 090423), by ALMA at $\nu$=222 GHz (Berger et al. 2014) and ATCA at
$\nu$=37.5 GHz (Stanway et al. 2011), respectively. We notice that the host luminosity of GRB 980425 is
about three orders of magnitude dimmer than the average host spectral luminosity $\sim3.6\times10^{29}$
erg s$^{-1}$Hz$^{-1}$ of all the measured host flux densities, although it is already the brightest radio host among these samples.
It is worth pointing out that at leat about 92 percent of these radio hosts
can be obtained by FAST and SKA successfully.

\subsection{Detection Rates}

As usual, one can calculate the GRB rates by assuming that GRBs and star formation rate (SFR) are
closely related so that GRBs trace the SFR exactly. Here we follow Y\"{u}ksel et al. (2008) to predict the
detection rates of GRBs by the current and future large radio instruments such as several upcoming
SKA pathfinders, FAST, Australian Square Kilometre Array Pathfinder (ASKAP), MeerKAT, etc. The number of GRBs detectable in the redshift range of
$z=0$~--~4 (Y\"{u}ksel et al. 2008) can be given by
\begin{equation}
\label{Detecting N0-4}
 \mathcal{N}_{0\rightarrow4}^{obs}=\Delta t\frac{\Delta \Omega}{4\pi}\int_0^4 dzF(z)\varepsilon(z)\dot{\rho_{*}}(z)\frac{dV(z)/dz}{1+z},
\end{equation}
where $\Delta t$ and $\Delta\Omega$ are the total live time and the angular sky coverage of the
telescope, respectively; $F(z)\equiv F_0$ and $\varepsilon (z)=\varepsilon_0(1+z)^\zeta$ have been
defined with two unknown constants ($F_0$ and $\varepsilon_0$) and $\zeta\simeq1.5$ has been
taken by Kistler, et al. (2008); $1/(1+z)$ is the correction factor due to cosmological time dilation; $dV(z)/dz=4\pi(c/H_0)D^{2}_{c}(z)/\sqrt{(1+z)^3\Omega_m+\Omega_\lambda}$ represents the comoving volume
per unit redshift where the comoving distance $D_c(z)$ is related with the luminosity distance
$D_l(z)$ by $D_l(z)=(1+z)D_c(z)$; $\dot{\rho_{*}}(z)$ is the star formation rate function which is
usually assumed as (Hopkins \& Beacom 2006),
\begin{equation}
\dot{\rho_{*}}(z)=\dot{\rho}_0\left[(1+z)^{a\eta}+(\frac{1+z}{B})^{b\eta}+(\frac{1+z}{C})^{c\eta}\right]^{1/\eta},
\end{equation}
with $a=3.4, b=-0.3, c=-3.5$, $\dot{\rho}_0=0.02 M_{\odot}$ yr$^{-1}$Mpc$^{-3}$, $\eta\simeq-10$,
$B\simeq5000$ and $C\simeq9$ (Y\"{u}ksel et al. 2008). Then the comoving event rate of GRBs can be calculated from $\dot{n}_{GRB}(z)=\varepsilon(z)\dot{\rho_{*}}(z)$.

Using Eq. (10), we can estimate the all-sky number of detectable GRBs up to a certain redshift $z$ as
\begin{equation}
\label{Detecting N0-z}
 \mathcal{N}_{0\rightarrow z}^{obs}=\mathcal{N}_{0\rightarrow4}^{obs}\times\frac{\Delta \Omega_1\Delta t_1}{\Delta \Omega\Delta t}\frac{\int_0^z dz(1+z)^{\alpha-1}\dot{\rho_{*}}(z)dV(z)/dz}{\int_0^4 dz(1+z)^{\alpha-1}\dot{\rho_{*}}(z)dV(z)/dz},
\end{equation}
where $\Delta t_1$ and $\Delta\Omega_1$ stand for the total observation time and the
angular sky coverage of the telescope. The observed GRB number is mainly determined by the
observation time, the field of view (FoV), and the sensitivity. Especially, for a GRB to be
detected, the observed flux density should be above the instrumental flux threshold given by
$F_{th,\nu}=(1+z_{max})L_{\nu}[4\pi D_l^2(z_{max})]^{-1}$ (k-correction not included here),
where $L_{\nu}$ is the spectral luminosity at the observing frequency $\nu$ and
$z_{max}$ is the maximal observable redshift for the burst.
Note that the detection rate will slightly decrease if the k-correction effect is taken into account.
In Figure 10, we plot the peak spectral luminosity-redshift distribution for the observed radio afterglows.
The redshifts of these GRBs generally range from $z=0$ to 4.
From this plot, we obtain the mean peak luminosity of radio afterglows as
$4_{-1}^{+12}\times10^{30}$ erg s$^{-1}$Hz$^{-1}$.

%\begin{equation}
%\label{Detecting N0-4}
% \mathcal{N}_{0\rightarrow 4}^{obs}\equiv\mathcal{N}_{0\rightarrow 4}^{exp}\times{\langle f_b\rangle},
%\end{equation}
%and
%
%\begin{equation}
%\label{Detecting N0-4}
% \mathcal{N}_{0\rightarrow z}^{obs}\equiv\mathcal{N}_{0\rightarrow z}^{exp}\times{\langle f_b\rangle}.
%\end{equation}

We have applied Eq. (12) to calculate the detection rate of radio afterglows versus the threshold
flux at ten typical frequencies. The results are illustrated in Figure 11. We find that FAST is more
powerful than most other existing or upcoming instruments, except for SKA (see Table 3). For example,
FAST has a theoretical sensitivity of 2 $\mu$Jy at $\nu=$1.4 GHz for an integration time of 1 hour, which is much better than the other two SKA pathfinders, i.e., MeerKAT with 30 $\mu$Jy and ASKAP with 60 $\mu$Jy (it is also noticeable that the upgraded Giant Meterwave Radio Telescope (uGMRT), as one of pathfinders of the SKA, works in 1420-150 MHz bands with a few hundred MHz bandwidth and can reach sensitivity up to 10-20 $\mu$Jy in various bands within a few hours of integration). It is capable of detecting
$\sim270$ GRB radio afterglows per square degree per year at $\nu=$1.4 GHz. The detection rate is thus higher than VLA by about
one order of magnitude. SKA is expected to acquire an even better sensitivity of 0.5 $\mu$Jy in
reality, and it will then generate an even higher detection rate of 464 deg$^{-2}$yr$^{-1}$ at the same
frequency. But it should also be noted that we have neglected two observational effects in our
calculations, i.e. the ``confusion'' effect and the ``baseline drift'' effect. These effects
generally would cause the wide band (i.e., continuum) observations at frequency $\nu<$ 5GHz much
more difficult for a single dish radio telescope (Condon 2002). Firstly, the confusion noise
will not go down even if we increase the integration time. Thanks to the broader FoV, huge single
dishes can image relatively large areas and smooth those low-brightness sources to complement
interferometric observations. In other words, interferometers including the JVLA may run rings
around arecibo-like single dishes for GRB continuum studies unless the above primary problems
are successfully solved technically (see also Chandra 2016; Chandra et al. 2016). The second serious problem
for the single dish will be baseline drifts caused by small receiver gain fluctuations
and by changing spillover as the galaxy is tracked. These baseline drifts can be mitigated by various scanning and beam-switching
schemes, but they are very inefficient and will occupy a lot of telescope time (private communications with Prof. D. A. Frail 2015). In addition, all kinds of Radio Frequency Interferences (RFI) around may also play un-negligible role on the single dish receivers. These deeply motivate us to consider how to overcome these similar puzzles for FAST. Hopefully, our results can shed new light on the studies of radio afterglows and hosts with the next-generation large telescopes, but need more technical developments to solve the above problems for the single-dish observations.

\section{Conclusions}

Based on the currently available radio data set, we analyze the statistical properties of
GRB afterglows and hosts, paying special attention to the flux-redshift dependence of both
afterglows and hosts. We have also investigated the detectability of GRB afterglows and
host galaxies at very high redshifts by different large radio telescopes. Our results
are summarized as follows.

\begin{itemize}
\item We verify the prediction that the observability of GRBs is largely independent
    of redshifts. Theoretically, we show that this feature is expected in
    the standard forward shock model for a thin shell expanding in either an ISM and/or a wind
    environment. When comparing with the observational data points, however, it is hard to
    distinguish which medium model is better since many of the microphysical parameters could
    vary at a certain range. Particularly, the fourth power law relation of $n\propto(1+z)^4$ is ruled out
    based on current observations, which is consistent with previous work of Gou et al. (2004).
\item Using our samples of radio hosts, we have investigated the dependence of the host
    flux density on the cosmological redshift. A trend that the radio host flux becomes
    less dependent on the redshift at farther distances is found, which implies the
    detectability of radio hosts may also be largely unrelated with redshift. Assuming
    a power law spectrum of $F_{\nu,h}\propto \nu^{\beta_{h}}$ for inspecting the corresponding flux-redshift relation, we have used the observed host flux densities to constrain the spectral index of $\beta_h$ ranging from -1 to 2.5 for most host galaxies.
  %  More interestingly, we found that radio hosts, like GRBs themselves, could be intrinsically classified into
%    two groups, radio-bright ones with $\beta_h\sim2$ and radio-faint ones with $\beta_h<1$.
    This may impose strong constraints on the GRB physics and galaxy evolution theories.
    However, the radio spectral index of GRB host galaxies is only deduced from a
    limited number of events and needs to be confirmed by more samples in the era of larger telescopes.
\item Finally, we have explored the detection rates of GRB afterglows by different
    large radio telescopes such as FAST, LOFAR, MeerKAT, ASKAP and SKA. FAST has an
    outstanding potential for very high redshift radio afterglows. Therefore, we stress
    that if FAST as a single dish telescope can overcome the so-called ``confusion and
    baseline drift'' difficulties for continuum observations at lower frequency of
    $\nu<$ 5 GHz, it would be able to detect a large number of radio afterglows and thus
    play an important role in detecting these faint radio sources in the near future.
    Optimistically, FAST is expected to be better than other SKA pathfinders at
    higher frequency, say $\nu>3$ GHz, hopefully in its second phase.
\end{itemize}

\acknowledgments
We would like to thank the anonymous referee for constructive suggestions and
valuable comments that led to the overall improvement of this study.
We thank Dale A. Frail, Bing Zhang, and Xuefeng Wu for their helpful comments and
discussions. We also appreciate Rick Perley and Daniel Perley for discussing the host
observations with VLA. This work is partly supported by the National Key R \& D Program of China (No. 2017YFA0402600), and the National Natural Science Foundation of China (Grant numbers: U1431126, 11873030, 11473012, 11690024, 11725313) and the CAS International Partnership Program (No.114A11KYSB20160008). P.C. acknowledges support from the Department of Science and Technology via SwarnaJayanti Fellowship Award (File no. DST/SJF/ PSA-01/2014-15). Z.B.Z. acknowledges the warm hospitalities
of Department of Physics \& Astronomy at University of Nevada Las Vegas, and National Radio
Astronomy Observatory at Socorro, New Mexico, USA.
 \begin{figure}
   \centering
  \includegraphics[angle=0,width=120mm,height=160mm]{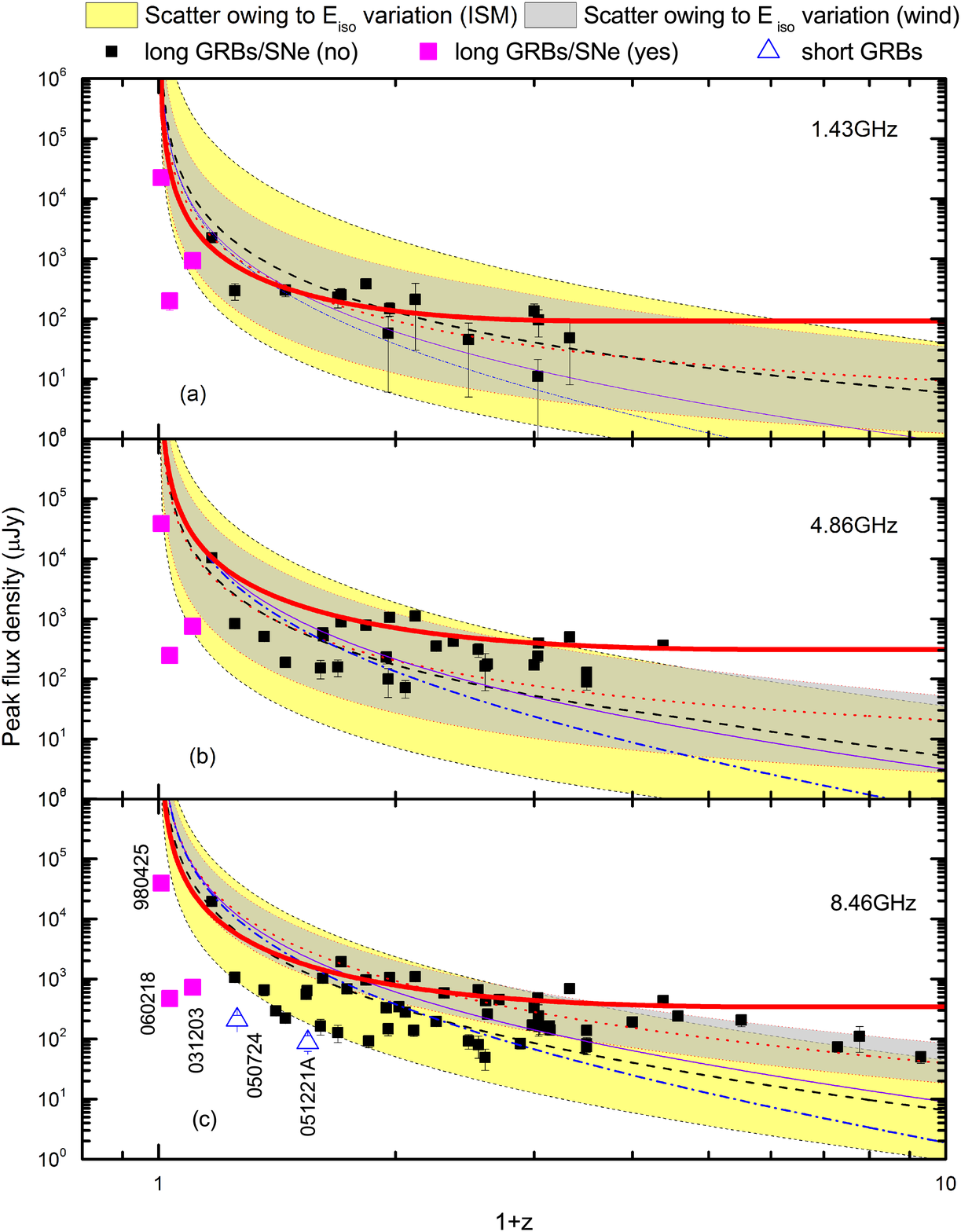}
      \caption{Peak flux density versus redshift for GRB radio afterglows, with the effects
      of the parameter $E_{iso}$ being specially illustrated.
      Panels (a) -- (c) correspond to GRBs at the observing frequency of 1.43 GHz, 4.86 GHz,
      and 8.46 GHz, respectively. The symbols have been marked on the legend. The dash-dotted line is plotted with the flux density
      evolving according to the inverse square of the luminosity distance. The solid line is the
      flux density scaling with an additional negative k-correction effect (see text). The light yellow
      and gray regions represent the flux-redshift dependencies in the scenarios of ISM and wind cases, respectively, for an average isotropic energy of $E_{iso}=2\times10^{52}$ erg and with one order of magnitude scatter. The thick solid line represents
      the fourth power-law relation of $n=n_0(1+z)^4$ with $n_0=0.1$ cm$^{-3}$. }
         \label{flux-z of radio afterglows}
   \end{figure}
\clearpage
 \begin{figure}
   \centering
  \includegraphics[angle=0,width=120mm,height=160mm]{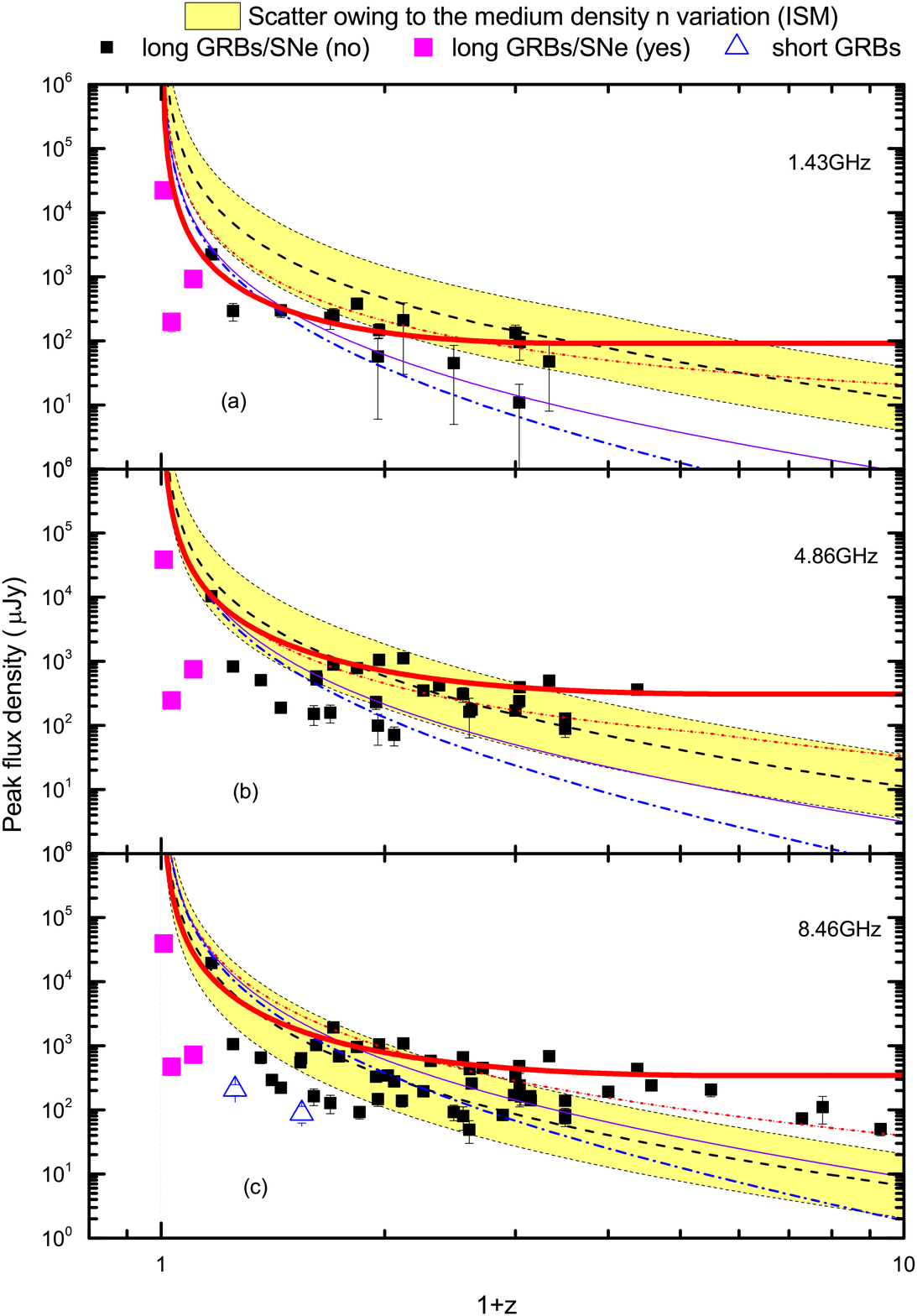}
      \caption{Peak flux density versus redshift of GRB radio afterglows, with the effects of the parameter $n$ being
      specially illustrated. The light yellow regions represent the flux-redshift dependencies in
      the scenarios of homogeneous ISM case, for an average interstellar medium density of
      $n=0.1$ cm$^{-3}$ and with one order of magnitude scatter. All other symbols are the same as in Figure 1. }
         \label{flux-z of radio afterglows}
   \end{figure}
\clearpage

%\clearpage
 \begin{figure}
   \centering
  \includegraphics[angle=0,width=120mm,height=160mm]{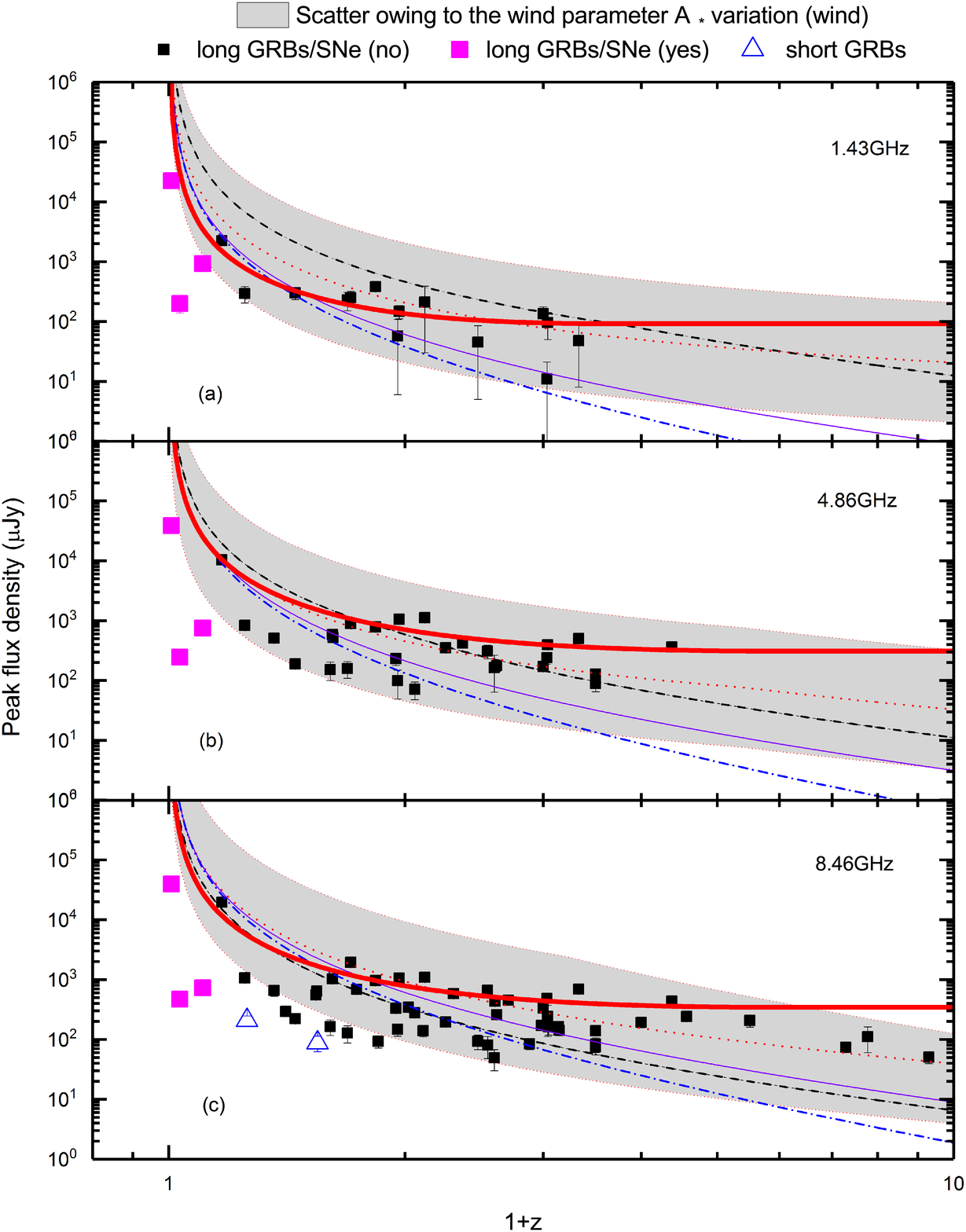}
      \caption{Peak flux density versus redshift of GRB radio afterglows, with the
       effects of the parameter $A_*$ being specially illustrated. The light gray regions
       represent the flux-redshift dependencies in the scenarios of the wind case for an average wind parameter of
     $A_*=0.2$ and with one order of magnitude scatter. All other symbols are the same as in Figure 1. }
         \label{flux-z of radio afterglows}
   \end{figure}
\clearpage

%\clearpage
 \begin{figure}
   \centering
  \includegraphics[angle=0,width=120mm,height=160mm]{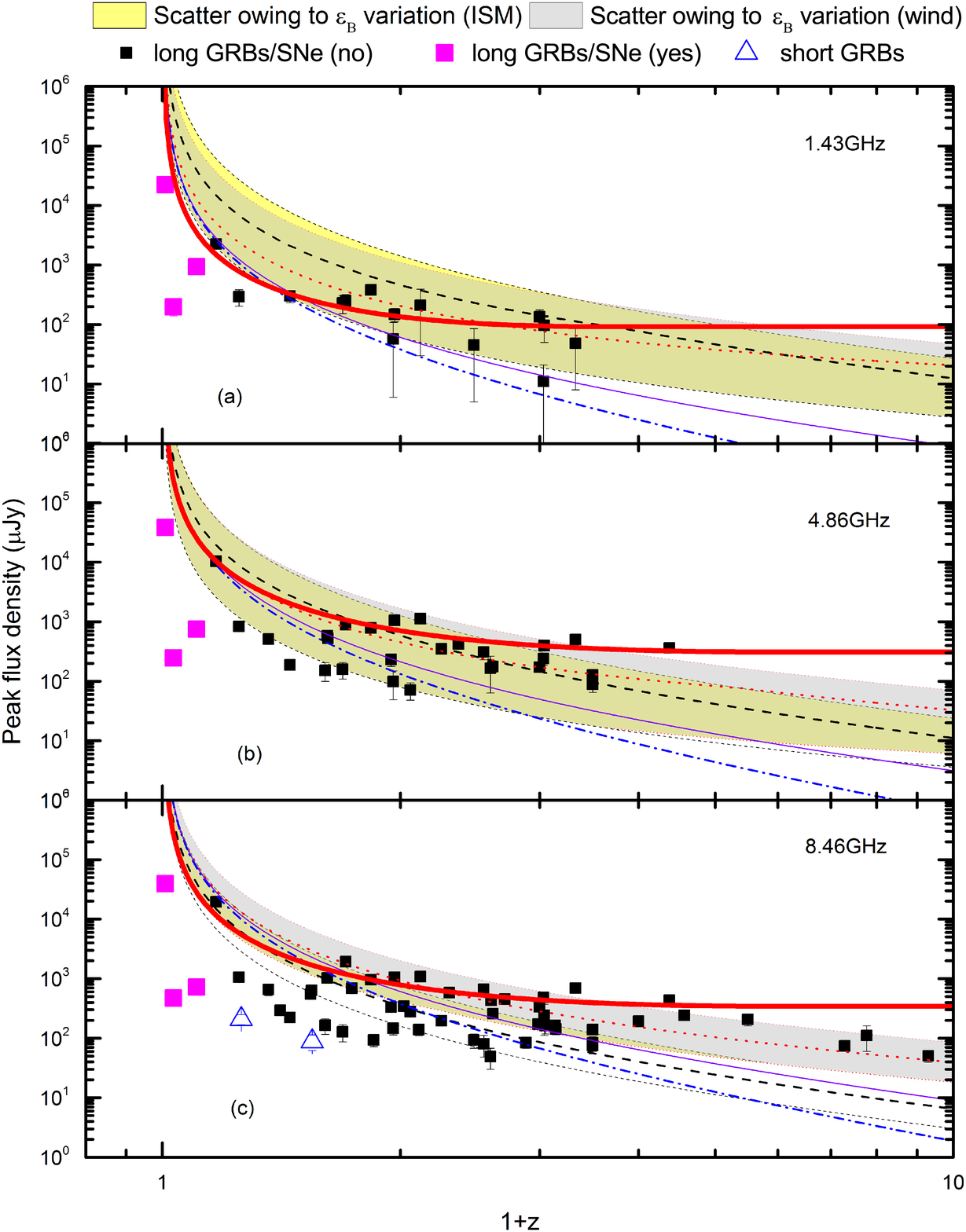}
      \caption{Peak flux density versus redshift of GRB radio afterglows, with the
      effects of the parameter $\varepsilon_{B}$ being specially illustrated. The light yellow
      and gray regions represent the flux-redshift dependencies in
      the scenarios of the ISM and wind cases, respectively, for an average magnetic field of
      $\varepsilon_B$=0.01 and with one order of magnitude scatter. All other symbols are the same as in Figure 1. }
         \label{flux-z of radio afterglows}
   \end{figure}
\clearpage

%\clearpage
 \begin{figure}
   \centering
  \includegraphics[angle=0,width=120mm,height=160mm]{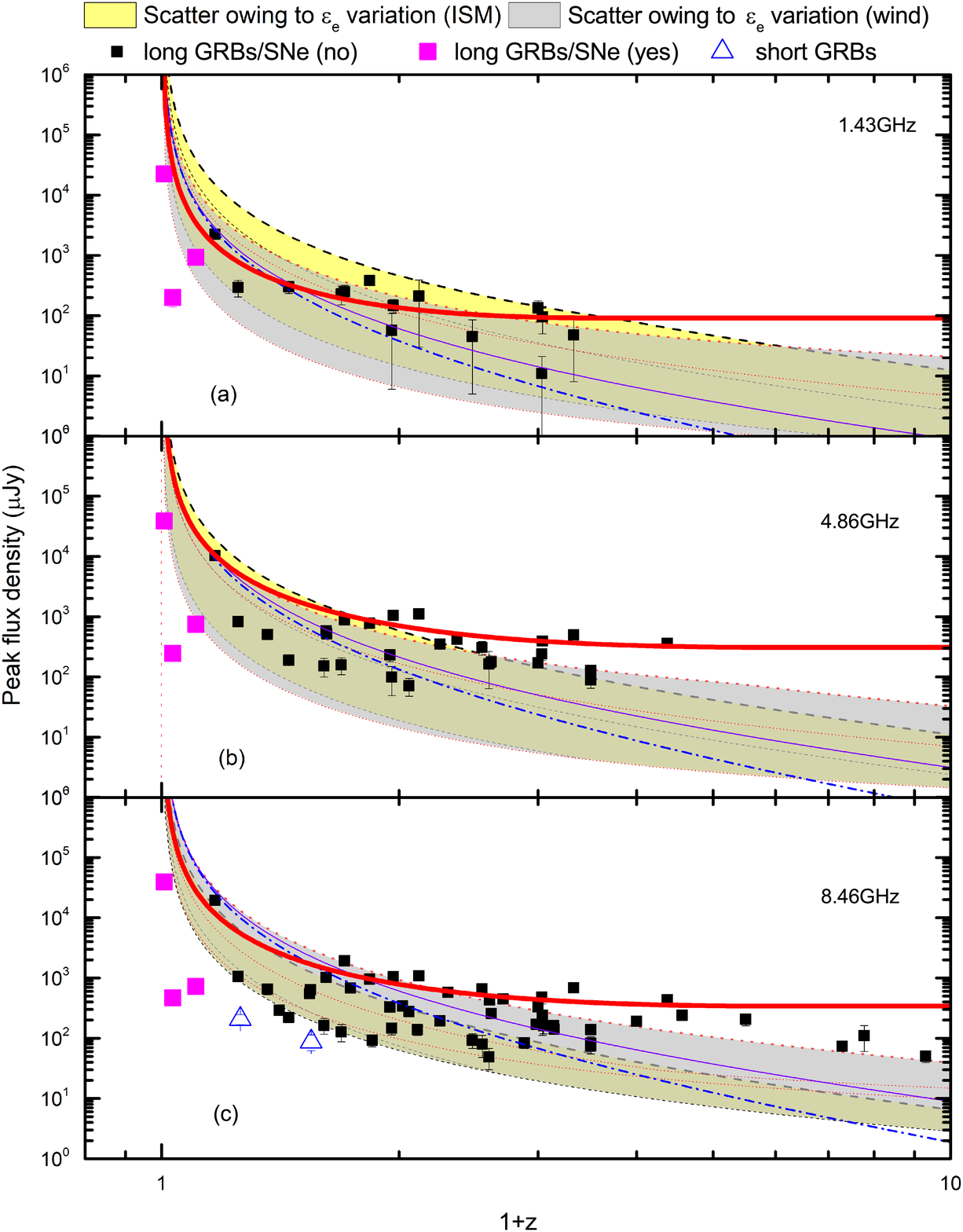}
      \caption{Peak flux density versus redshift of GRB radio afterglows, with the effects of
      the parameter $\varepsilon_e$ being specially illustrated. The light yellow
      and gray regions represent the flux-redshift dependencies in the scenarios of the ISM and wind cases,
      respectively, for an average electron parameter of
      $\varepsilon_e$=0.1 and with one order of magnitude scatter. All other symbols are the same as in Figure 1.}
         \label{flux-z of radio afterglows}
   \end{figure}
\clearpage

 \begin{figure}
   \centering
\includegraphics[angle=0,scale=0.5]{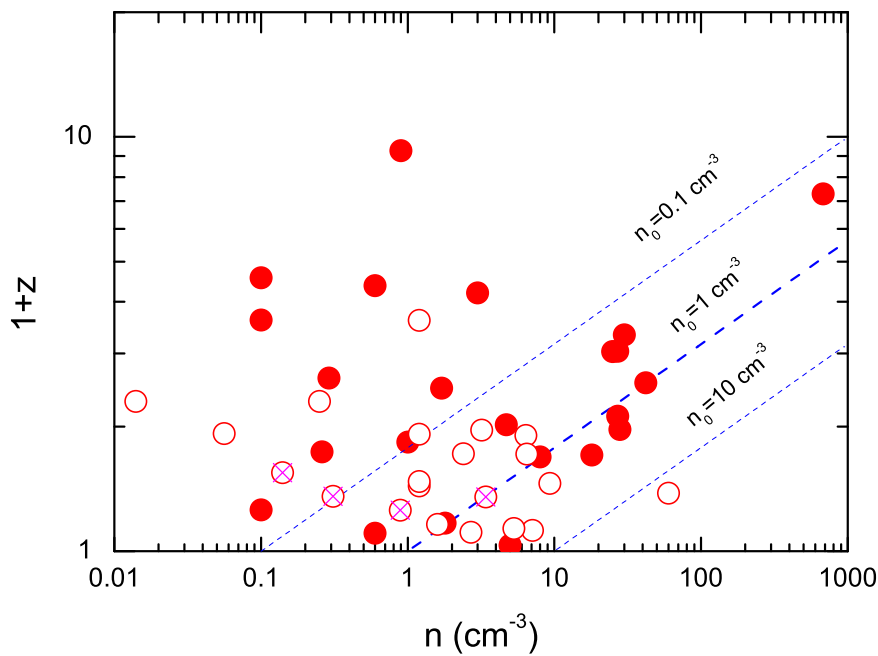}
      \caption{Theoretically derived medium density versus redshift for a number of GRBs.
      The data are mainly taken from Chandra \& Frail (2012) and Fong et al (2015).
      The solid and empty circles represent 24 long and 21 short GRBs, respectively.
      The four cross-circles stand for short bursts with radio afterglows detected so far.
      The dashed lines show the different density forms of $n=n_0 (1+z)^4$ with $n_0$=10, 1 and 0.1 cm$^{-3}$, respectively.}
         \label{n-z relation}
   \end{figure}
%\clearpage
%
 \begin{figure}
   \centering
   \includegraphics[angle=0,width=160mm,height=80mm]{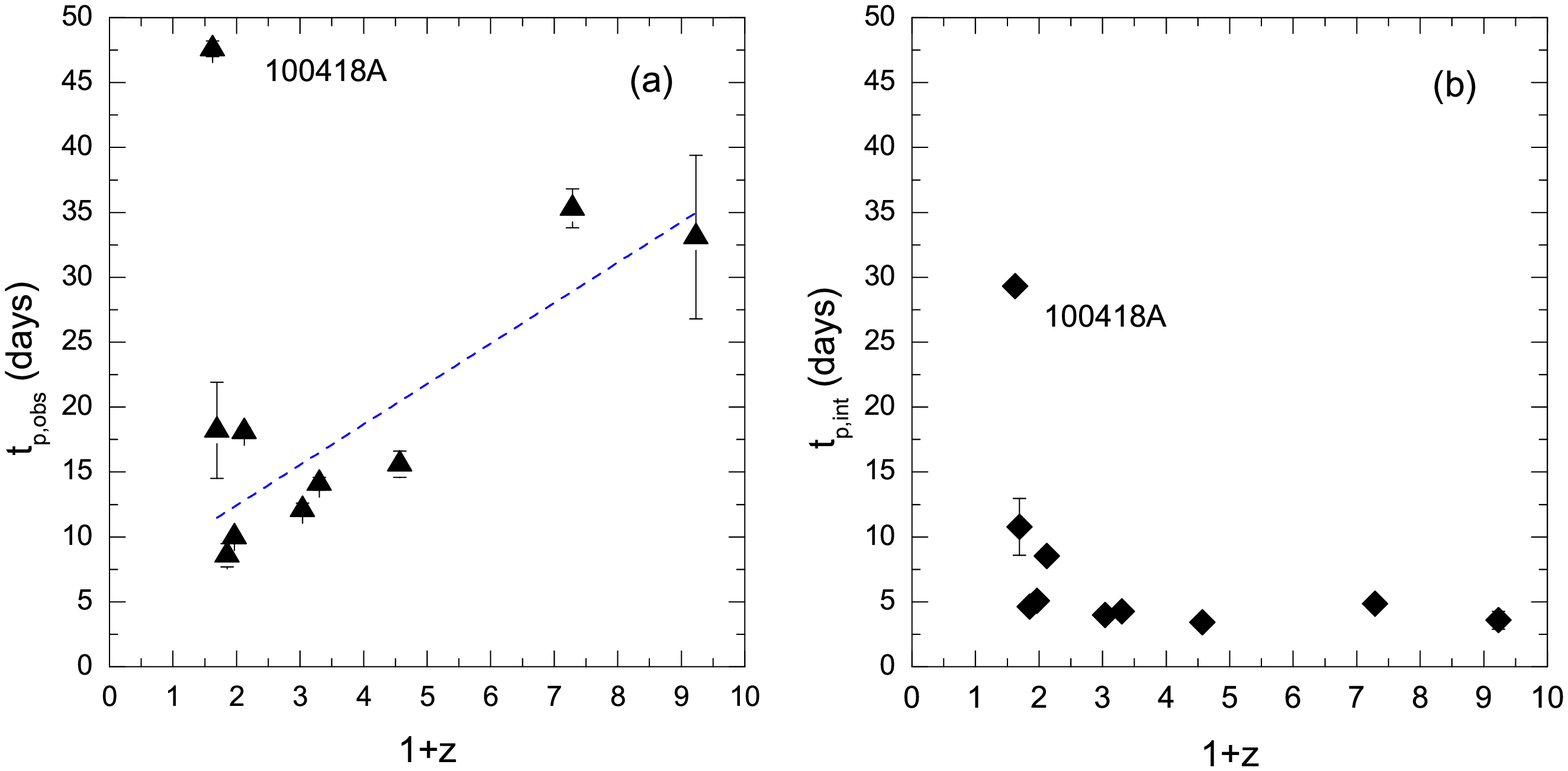}
      \caption{Correlation between the redshift and the peak time of 8.5 GHz radio afterglows.
      In Panel (a), the Y-axis is simply the observed peak time, while in Panel (b) the Y-axis
      is the intrinsic peak time (i.e., corrected for the cosmological time dilation effect).
      Note that GRB 100418A seems to be an outlier in these plots, the reason of which is still quite uncertain.
      The best linear fit to the nine bursts except GRB 100418A is shown by the dashed line in Panel (a).}
         \label{peak times vs. redshift. fig3}
   \end{figure}

%\clearpage
 \begin{figure}
  \centering
 \includegraphics[angle=0,scale=0.5]{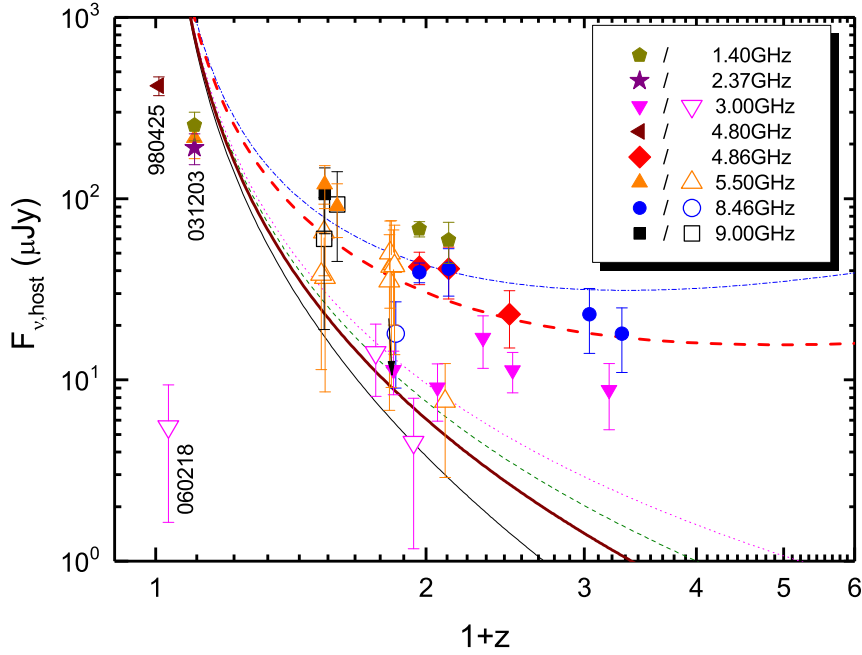}
      \caption{Host radio flux vs. redshift at multiple frequencies for the observational data
      in Table 1 (not including the upper limits). The thin solid line corresponds to the simple
      inverse square law of the luminosity distance without K-correction ($\beta_h=-1$);
      The remaining lines represent different scenarios for K-corrections [thick solid line: $\beta_h=-1/3$ (Berger, Kulkarni \& Frail 2001); dashed: $\beta_h=0$; dotted: $\beta_h=1/3$; red thick dashed: $\beta_h=2$; dash-dotted: $\beta_h=2.5$]. Observational data points at different frequencies are denoted by the diverse solid/empty symbols for larger/smaller than $3\sigma$ confidence levels, correspondingly.}
         \label{hostflux vs. z}
   \end{figure}

 %\begin{figure}
%  \centering
% \includegraphics[angle=0,scale=0.5]{f5.eps}
%      \caption{Host radio flux vs. redshift for the whole sample of radio hosts in table 1 (N=37). All symbols are the same as in Fig. 4.}
%         \label{hostflux vs. z}
%   \end{figure}

   \begin{figure}
  \centering
 \includegraphics[angle=0,scale=0.45]{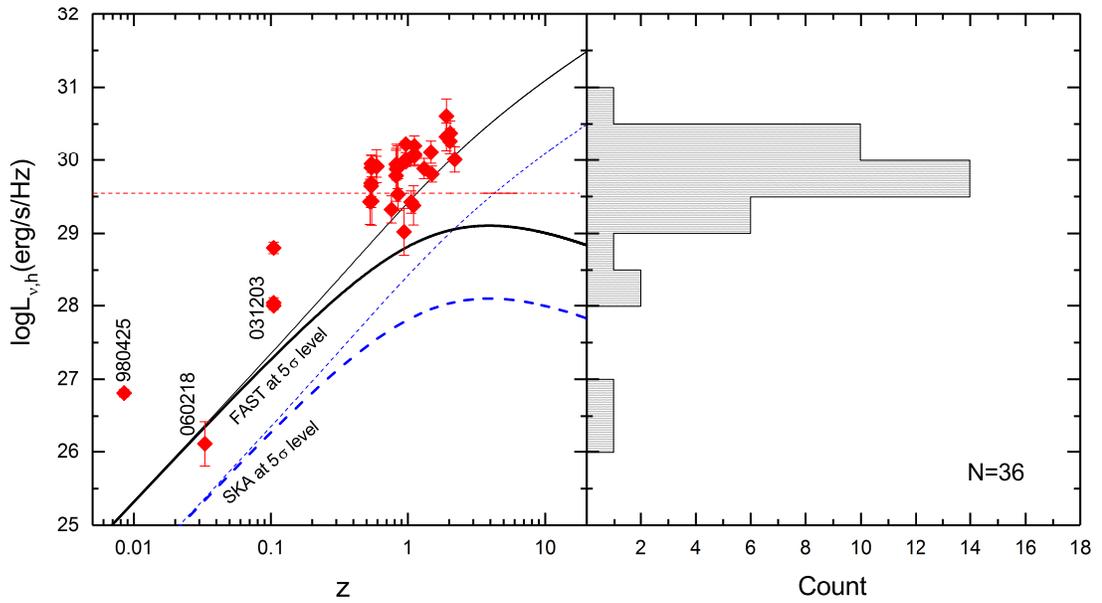}
      \caption{Radio spectral luminosity versus redshift (left panel) and luminosity histogram of 36 host
      flux densities (right panel).
      % The horizontal dot line indicates the averaged radio luminosity ($\sim3.6\times10^{29} $ erg s$^{-1}$Hz$^{-1}$) of host galaxies.
       The sensitivities of FAST and SKA with (thick curves) and without
       (thin curves) k-correction are given for $\nu=1.43$ GHz at $5\sigma$ level by assuming a 1 hr integration time.}
         \label{HostLnu-z}
   \end{figure}

   \begin{figure}
   \centering
 \includegraphics[angle=0,scale=0.6]{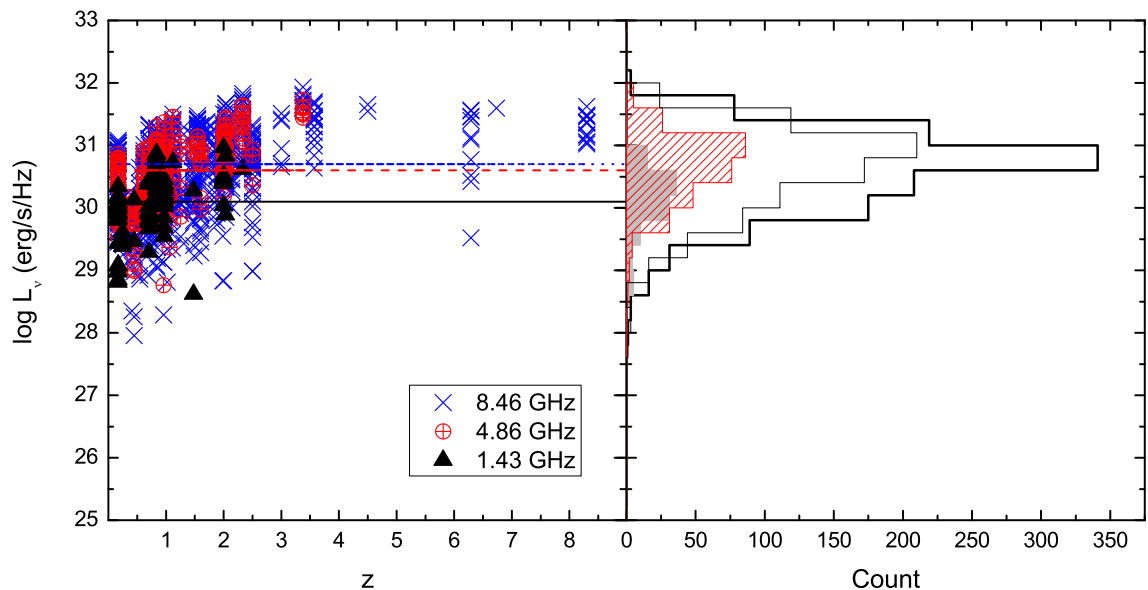}
      \caption{Left panel: peak luminosities versus redshifts for GRB radio afterglows,
      with 101, 279 and 784 measurements at 1.43 GHz, 4.86 GHz and 8.46 GHz, respectively.
      The corresponding average spectral luminosities are denoted by three horizontal lines,
      which are in the range of $1\times10^{30}$ --- $5\times10^{30}$ erg/s/Hz.
      Right panel: radio luminosity distributions for the 1.43 GHz (shade), 4.86 GHz (hatched),
      8.46 GHz (thin line) and the whole (thick line) samples.}
         \label{afterglowLnu-z}
   \end{figure}

\begin{figure}
   \centering
 \includegraphics[angle=0,scale=0.5]{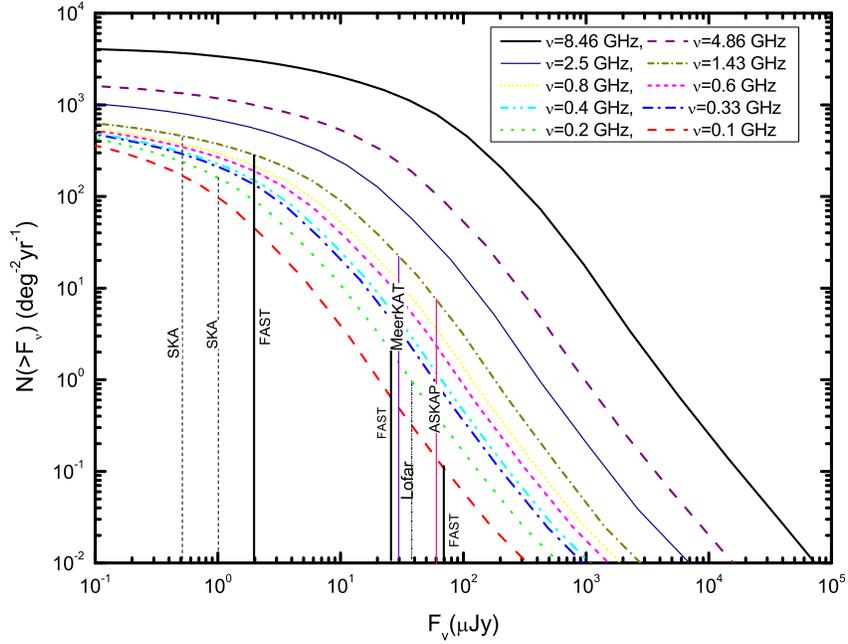}
      \caption{Cumulative flux distributions of radio afterglows at various observational
      frequencies. The vertical lines indicate the detection limits of different instruments,
      including LOFAR, ASKAP, FAST, MeerKAT, and SKA. The detecting sensitivities are
      calculated by assuming $\Delta \tau$ = 1 hr, $\Delta \nu$ = 100 MHz and S/N = 5.
      Note that the vertical lines in this figure only refer to the sensitivity of the
      instrument at the frequency located at the top of the line.
 }
         \label{gamma-theta}
   \end{figure}
 %\begin{figure}
%   \centering
% %\includegraphics[angle=0,scale=1.5]{fig4.eps}
%      \caption{The Lorentz factor $\gamma$ vs. the viewing angle $\theta$. \textit{Solid line}
%   corresponds to $\delta=1$; \textit{dashed line} represents $\gamma=1/sin\theta$; and \textit{dotted
%   line} describes $\beta_{app}=1$.}
%         \label{gamma-theta}
%   \end{figure}
%% If you are not including electonic art with your submission, you may
%% mark up your captions using the \figcaption command. See the
%% User Guide for details.
%%
%% No more than seven \figcaption commands are allowed per page,
%% so if you have more than seven captions, insert a \clearpage
%% after every seventh one.

%% Tables should be submitted one per page, so put a \clearpage before
%% each one.

%% Two options are available to the author for producing tables:  the
%% deluxetable environment provided by the AASTeX package or the LaTeX
%% table environment.  Use of deluxetable is preferred.
%%

%% Three table samples follow, two marked up in the deluxetable environment,
%% one marked up as a LaTeX table.

%% In this first example, note that the \tabletypesize{}
%% command has been used to reduce the font size of the table.
%% We also use the \rotate command to rotate the table to
%% landscape orientation since it is very wide even at the
%% reduced font size.
%%
%% Note also that the \label command needs to be placed
%% inside the \tablecaption.

%% This table also includes a table comment indicating that the full
%% version will be available in machine-readable format in the electronic
%% edition.
%
\clearpage

\begin{deluxetable}{lccccccccc}
\tabletypesize{\scriptsize}
%\rotate
\tablecaption{Observed Parameters of Radio Host Galaxies of GRBs.}
\tablewidth{0pt}
\tablehead{
%\colhead{GRB} & \colhead{$T_{90}$ } & \colhead{z} & \colhead{$E_{\gamma,iso}$}&$\nu _{obs}$ & \colhead{$F_{host}$} &
% $F_{peak}$ & \colhead{$t_{peak}$}  & \colhead{Refs}& Telescope\\
%&(s)&&($10^{51}$ erg)&(GHz)&($\mu$Jy)&($\mu$Jy)&(days) &&\\
%(1)&(2)&(3)&(4)&(5)&(6)&(7)&(8) &(9)&(10)
\colhead{GRB} & \colhead{$T_{90}$ } & \colhead{z} & \colhead{$E_{\gamma,iso}$}&$\nu _{obs}$ & \colhead{$F_{host}$}  & \colhead{Refs}& Telescope\\
&(s)&&($10^{51}$ erg)&(GHz)&($\mu$Jy)& & \\
(1)&(2)&(3)&(4)&(5)&(6)&(7)&(8)
}\startdata
980425$^{\ddag}$& 31& 0.0085&0.002&4.8  & 420$\pm$50$^{\star}$ &1& ATCA \\
                &   &       &     &8.64  &$<180$              &1& ATCA \\
980703& 90& 0.966&69&1.43  & 68$\pm$6.6$^{\star}$  &    2&VLA\\
      &   &   &     &4.86  & 42.1$\pm$8.6$^{\star}$ &  2&VLA \\
      &   &   &     &8.46  & 39.3$\pm$4.9$^{\star}$ &  2&VLA\\
000210& 10& 0.85&200&8.46  &  18$\pm$9   & 4&VLA\\
000301C& 10& 2.034&43.7&8.46  & 18$\pm$7$^{\star}$ &  3&VLA \\
000418& 30& 1.119&75.1&1.43  & 59$\pm$15$^{\star}$ &  3&VLA\\
      &   &      &    &4.86  & 41$\pm$13$^{\star}$ &  4&VLA \\
      &   &      &    &8.46  & 41$\pm$12$^{\star}$ &  4&VLA \\
000926& 25& 2.039&270&8.46  & 23$\pm$9$^{\star}$  &   3&VLA\\
010222& 170& 1.477&133&4.86  & 23$\pm$8$^{\star}$  & 3&VLA\\
011121$^{\ddag}$& 105& 0.362&45.5&4.86  & $<$120 &  12&VLA\\
020405$^{\ddag}$& 40& 0.69 &110 &8.46  & $<$42   &  9&VLA \\
031203$^{\ddag}$& 30& 0.105&0.115&1.39  & 254$\pm$46$^{\star}$  & 10&ATCA \\
                         &   &      &     &2.37  & 191$\pm$37$^{\star}$  & 10&ATCA \\
                          &   &      &     &5.5  & 216$\pm$50$^{\star}$  & 11&ATCA \\
050525A$^{\ddag}$& 9& 0.606& 20.4&5.5  & $<$15.6   & 5 &ATCA\\
050824$^{\ddag}$& 23& 0.83&  1.5 &5.5  & 42.3$\pm$33.2   & 5 &ATCA\\
%050904$^{\dag}$& 174& 6.29&1300&8.46  & 13$\pm$6 & 6&VLA \\
051022&200& 0.809& 630&5.5  & $<$23.0   &  5 &ATCA\\
060218$^{\ddag}$& 128& 0.033&0.003&3.0  & 5.52$\pm$3.88 & 8 &VLA\\
%090323$^{\dag}$& 133& 3.57&4100&8.46  & 27$\pm$14 &  6&VLA \\
%090423$^{\dag}$& 10.3& 8.23&110&8.46  & 5.6$\pm$1.2$^{\star}$  & 6 &VLA\\
090423$^{\dag}$& 10.3& 8.23&110&37.5  &  $<$9.3 &  6 &ATCA\\
                         &  &    &     &222   &  $<$33 &  7 &ALMA\\
090424& 50& 0.544&44.7&5.5  & 36.6$\pm$28 & 5 &ATCA\\
%100418A$^{\ddag}$& 7& 0.623&0.52&5.5  & 363$\pm$48$^{\star}$ & 5&ATCA \\
%                 & &     &    &9.0  & 199$\pm$57$^{\star}$  &  5&VLA \\
\hline
050223& 22.5 & 0.592 & 0.87  &  5.5  & 90.5$\pm$30.1$^{\star}$ &  5&ATCA\\
     &       &      &        &  9.0  & 93$\pm$48&  5&VLA\\
050922C& 4.5 &2.198 &  37.4  &  3.0  & 8.8$\pm$3.5$^{\star}$ &  8&VLA\\
051006& 34.8 &1.059 & 35.8  &  3.0  & 9.08$\pm$3.17$^{\star}$ &  8&VLA\\
060729$^{\ddag}$& 115.3& 0.54 & 13.8  &  5.5  & 65.4$\pm$27.8&  5&ATCA\\
             &       &      &       &  9.0  & 60$\pm$41&  5&VLA\\
060814& 145.3 & 1.92 & 307  &  3.0  & 11.34$\pm$3.1$^{\star}$ &  8&VLA\\
      &    &    &      &  5.5  & 43.6$\pm$23.5&  5&ATCA\\
060908& 19.3 &1.884 & 44  &  3.0  & $<$4.53   &  8&VLA\\
060912A& 5 &0.937 &  17.3 &  3.0  & 4.54$\pm$3.37&  8&VLA\\
061110A& 40.7 &0.758 & 13.2  &  3.0  & 14.2$\pm$6.08&  8&VLA\\
061121& 81.3 &1.314 &  272 &  3.0  & 17.07$\pm$5.47$^{\star}$ &  8&VLA\\
070129& 461 &2.34 & 26.9  &  3.0  & $<$4.92   &  8&VLA\\
070306& 210 &1.497 & 88  &  3.0  & 11.31$\pm$2.84$^{\star}$ &  8&VLA\\
070506& 4.3 &2.31 & 4.23  &  3.0  & $<$3.69   &  8&VLA\\
070508& 20.9 & 0.82 & 70  &  5.5  & 35.0$\pm$28.2&  5&ATCA\\
071112C& 15 & 0.823 & 5.3  &  5.5  & 50.1$\pm$25.2&   5&ATCA\\
080413B& 8 & 1.1 &  16.5 &  5.5  & 7.6$\pm$4.7&   5&ATCA\\
080710& 120 & 0.845 & 49.5  &  5.5  & 42.6$\pm$28.8&   5&ATCA\\
081007$^{\ddag}$& 10 & 0.529 & 0.16  &  5.5  & 38.1$\pm$26.7&  5&ATCA\\
100621A& 63.6& 0.542 & 43.5  &  5.5  & 120$\pm$32$^{\star}$ &   5&ATCA\\
      &     &        &       &  9.0  & 106$\pm$42$^{\star}$&   5&VLA\\
\hline

\enddata
%% Text for table notes should follow after the \enddata but before
%% the \end{deluxetable}. Make sure there is at least one \tablenotemark
%% in the table for each \tablenotetext.
\tablecomments{References are given for the host radio flux density:
1. Michalowski et al. (2009); 2. Berger, Kullarni \& Frail (2001); 3. Perley \& Perley (2013); 4. Berger et al. (2003a); 5. Stanway et al. (2014); 6. Stanway et al. (2011); 7. Berger et al. (2014); 8. Perley et al. 2015; 9. Berger et al. (2003b); 10. Michalowski et al. (2012); 11. Stanway et al. (2010); 12. Frail et al. 2003\\
%$^{\dag}$ High redshift bursts without direct host observations. These host flux densities are
%estimated from our newly-derived RRF method (Li et al. 2015) at low/medium observing
%frequency, say $\nu=$8.46 GHz.\\
$^{\ddag}$ SN-associated GRBs. \\
$^{\star}$ Host flux densities that are larger than $3 \sigma$ level.\\
$^{\dag}$ High redshift GRBs. 
%$^{\pounds}$ Radio afterglows peaking at 4.86 or 8.46 GHz (Chandra \& Frail 2012).
}

%\tablenotetext{a} {\ ``Long'' GRBs with short-hard properties (see the text in section 2).}

%\tablenotetext{\ddag}{\ }
\end{deluxetable}
%% If you use the table environment, please indicate horizontal rules using
%% \tableline, not \hline.
%% Do not put multiple tabular environments within a single table.
%% The optional \label should appear inside the \caption command.
%
\clearpage %%%

\begin{deluxetable}{cccc}
\tabletypesize{\scriptsize}
%\rotate
\tablecaption{Power-law Index $\tau$ of the Peak Flux-Redshift Relation.}
\tablewidth{0pt}
\tablehead{
\colhead{Medium} &\colhead{Density form}& \colhead{$\tau$ in case I} & \colhead{$\tau$ in case II} \\
}
\startdata
ISM&$n=1 cm^{-3}$&1&$(7p+3)/[2(p+4)]\simeq1.5$\\
ISM&$n=(1+z)^4 cm^{-3}$&3&$(3p+27)/[2(p+4)]\simeq2.7$\\
\hline
wind& $n=3\times10^{35}A_{\ast}R^{-2} cm^{-3}$ &3/2&$(6p+9)/[2(p+4)]\simeq1.8$\\
\hline
\enddata
%% Text for table notes should follow after the \enddata but before
%% the \end{deluxetable}. Make sure there is at least one \tablenotemark
%% in the table for each \tablenotetext.
\tablecomments{For further details, other parameters involved ($n, A_{\ast}, R$ and $p$) can be
found in Section 3. Note that $F_{\nu, t_p}\propto(1+z)^{\tau}D_l^{-2}(z)$ has been
defined in the main text. }

%\tablenotetext{\ddag}{\ }
\end{deluxetable}
\clearpage
\begin{deluxetable}{ccccccccc}
\tabletypesize{\scriptsize}
%\rotate
\tablecaption{Key Parameters of Current and Future Radio Telescopes.}
\tablewidth{0pt}
\tablehead{
\colhead{Telesope} & \colhead{Frequency} & \colhead{Bandpass} & \colhead{$\nu_{obs}$}& $A_{eff}/T_{sys}$ & $\Omega_{FoV}^{\dag}$ & $F_{lim}$& Detection Rate& Ref\\
&(MHz)&(MHz)&(MHz)&($m^2/K$)&(deg$^2$)& ($\mu$Jy)& (\#/deg$^{2}$/yr)&\\
}
\startdata
VLA&75-43000&1000&1430&100-200&0.22&50&11& 1\\
&&4000&4860&&0.02&20&311\\
&&4000&8460&&0.01&13&1703\\
\hline
FAST&70-3000&70&100&2000&0.4&71&0.1&2\\
&&140&200&&0.1&26&2.2&\\
&&280&400&&0.025&&\\
&&460&800&&0.006&&\\
%&&[320-334]&328&&&\\
%&&[550-640]&600&&&\\
&&570&1450&&0.002&2&270&\\
%&&[1230-1430]&1380&&&\\
&&1000&2500&&0.001&&\\
\hline
LOFAR&10-80  &3.66&60 &400&74.99&&&3\\
     &110-240&3.66&150&400&11.35&38&1&\\
\hline
ASKAP&700-1800&300&1400&$>85$&30&60&7.2&4\\
\hline
MeerKAT&500-2000&1500&1400&$>160$&1.1&30&21&5\\
\hline
MWA&80-300&30.72&150&7&610&&&6\\
\hline
SKA&50-20000&230&150&5000-10000&200&1&156&7\\
&&9700&700&&1-200&0.5&464&\\
&&10000&5500&&1&&&\\
\enddata
%% Text for table notes should follow after the \enddata but before
%% the \end{deluxetable}. Make sure there is at least one \tablenotemark
%% in the table for each \tablenotetext.
\tablecomments{References: 1. Thompson et al. 1980; 2. Nan et al. 2011; 3. van Haarlem et al. 2013; 4. Johnston et al. 2008; 5. Booth et al. 2009; 6. Tingay et al. 2013; 7. Dewdney et al. 2009.\\
$\dag$ The sky coverage is given by $\Omega =\pi(FoV/2)^2$, where the Filed of View ($FoV$) of a given telescope or array can be estimated with $FoV=1.22\times\frac{\lambda}{D}$, in which $\lambda$ is the observing wavelength and $D$ is the effective aperture or the maximal length of baseline between each dish pairs. For VLA, the $FoV$ is determined by $FoV=\frac{45}{\nu_{(GHz)}}$ arcmin. For FAST, we have $FoV=\frac{14}{\nu_{(GHz)}}$ arcmin at different frequencies with a constant $D=300$ m for the beam. All others are taken from the above references directly.}

%\tablenotetext{\ddag}{\ }
\end{deluxetable}
%% If you use the table environment, please indicate horizontal rules using
%% \tableline, not \hline.
%% Do not put multiple tabular environments within a single table.
%% The optional \label should appear inside the \caption command.
%
\clearpage %%%%

\end{document}